\documentclass[%
 reprint,
 amsmath,amssymb,
 aps,
]{revtex4-1}

\usepackage{graphicx}
\usepackage{dcolumn}
\usepackage{bm}
\usepackage[utf8x]{inputenc} 

\usepackage[caption = false]{subfig}
\usepackage{float}
\usepackage{subfig}

\graphicspath{ {./picture/} }

\begin{document}
\preprint{APS/123-QED}

\title{Effect of the charge distribution of virus coat proteins on the length of packaged RNAs}
\author{Yinan Dong}
\affiliation{Department of Physics and Astronomy,
University of California, Riverside, California 92521, USA}
\author{Siyu Li}\thanks{Present address:  Department of Materials Science and Engineering, Northwestern University,
Evanston, United States}
\author{Roya Zandi}
\affiliation{Department of Physics and Astronomy,
   University of California, Riverside, California 92521, USA}

\date{\today}

\begin{abstract}

Single-stranded RNA viruses efficiently encapsulate their genome into a protein shell called the capsid. Electrostatic interactions between the positive charges in the capsid protein’s N-terminal tail and the negatively charged genome have been postulated as the main driving force for virus assembly. Recent experimental results indicate that the N-terminal tail with the same number of charges and same lengths packages different amounts of RNA, which reveals that electrostatics alone cannot explain all the observed outcomes of the RNA self-assembly experiments. Using a mean-field theory, we show that the combined effect of genome configurational entropy and electrostatics can explain to some extent the amount of packaged RNA with mutant proteins where the location and number of charges on the tails are altered. Understanding the factors contributing to the virus assembly could promote the attempt to block viral infections or to build capsids for gene therapy applications.

\end{abstract}

\pacs{Valid PACS appear here}
\maketitle


\section{\label{sec:level1-1}Introduction}

\begin{figure}[t]
  \centering
\includegraphics[width=0.95\linewidth]{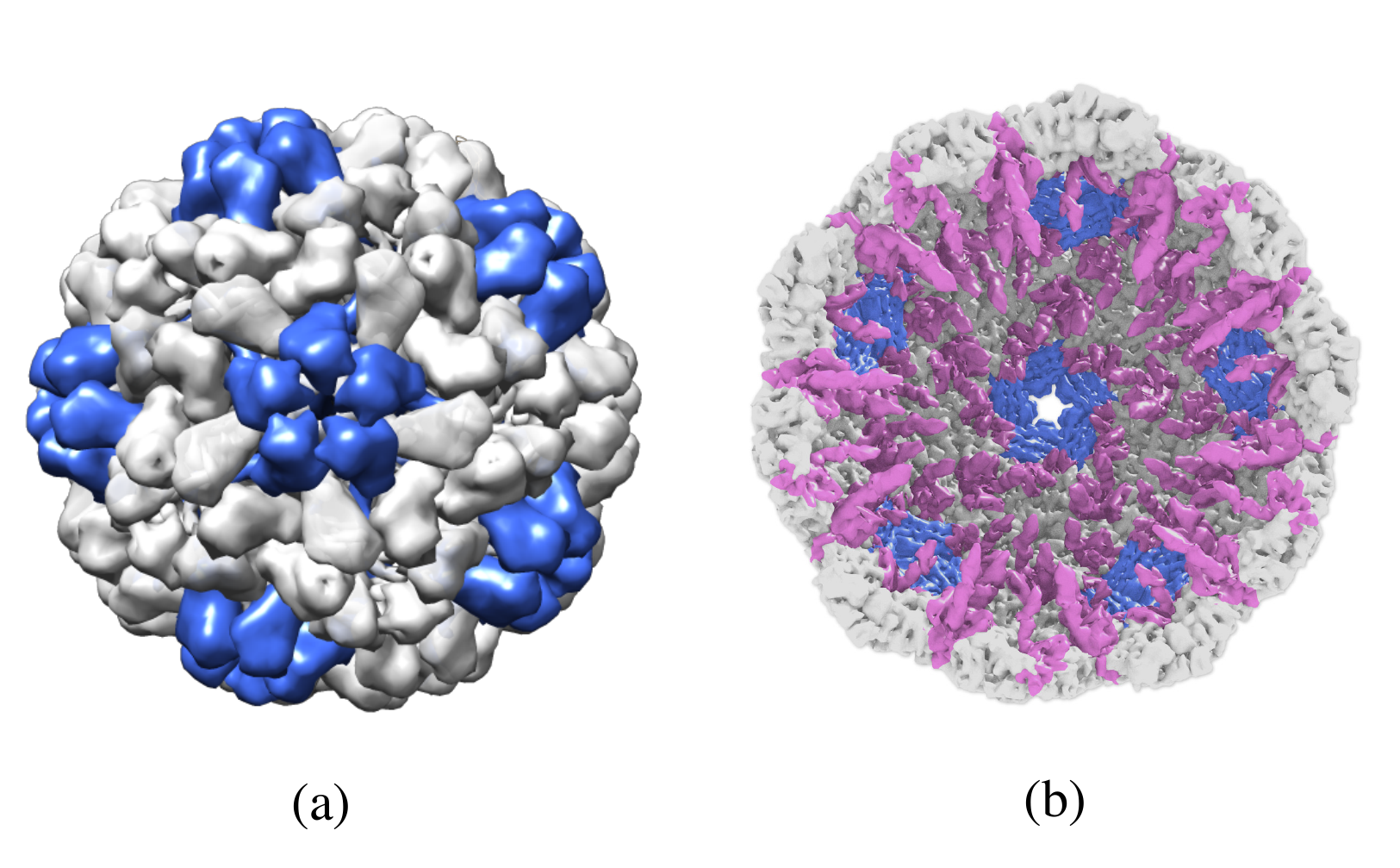}
  \caption{(a) A $T=3$ icosahedral shell with 180 protein subunits. The darker (blue) color shows the pentamers. The structure is similar to the BMV capsid. (b)  The interior of a $T=3$ viral shell with N-terminal domains (pink tails) extended toward the center of the capsid. Each N-terminal domain contains eight positive charges, not shown in the figure. The structure in (a) is reproduced using UCSF Chimera packages (http://www.rbvi.ucsf.edu/chimera).}
  \label{fig:BMV_inside}
\end{figure}

Viruses have optimized the feat of packaging of their negatively charged genomes into a protein shell called the capsid, often built from a large number of one or a few different kinds of protein subunits \cite{Bancroft}.  Under many {\it in vitro} conditions, coat proteins of several single-stranded RNA (ssRNA) viruses can spontaneously encapsulate all types of anionic cargos including their native genome, linear polymers, and heterologous and nonviral RNAs \cite{Comas,Beren2017,BORODAVKA2016,Zandi2016,nature2016}. The capsid proteins of several RNA viruses contain an unstructured positively charged N-terminal domain that extends toward the center of the capsid and interacts with the viral genome; see Fig.~\ref{fig:BMV_inside}\cite{Cadena2011}. Although the specific sequence of the viral RNA plays an important role in packaging \cite{Haganpackaging,Stockley2013}, it is now well established that the electrostatic interaction between N-terminal tails and RNA is the main driving force for the formation of viral particles and their stability \cite{Sun2007,Li2017,Zandi2020}. 

 Self-assembly studies of various ssRNA viruses have revealed that the amount of RNA packaged depends directly on the number of positive charges on the N-terminal tails of capsid proteins. Many experiments show that mutant virions with less positive charges on N-terminal domain encapsidate lower amounts of RNA and mutants with increased positive charges package more \cite{Venky2016,Bogdan}.  For example, the experimental studies of Sivanandam {\it et al.} show that the deletions of even one single positively charged residue of the satellite tobacco mosaic virus N-terminal domain results in the formation of virus particles with a reduced amount of viral RNAs \cite{Venky2016}. Belyi and Muthukumar as well as Hu {\it et al.} \cite{Belyi2006,Shklovskii} also examined the relation between the total number of positive charges in the tails and the length of the encapsidated RNA in various viruses and found a strong relation between them. 

Of particular interest is the self-assembly experiments of Ni {\it et al.} who specifically focused on the brome mosaic virus (BMV) and systematically investigated the role of electrostatics on the amount of RNA packaged \cite{Bogdan}. The N-terminal domain of BMV capsid proteins is composed of 26 residues, eight of which are positively charged.  The genome of BMV consists of four RNA molecules: RNA1 (3.2 kb), RNA2 (2.9 kb), RNA3 (2.1 kb), and RNA4 (0.9 kb). While RNA3 and RNA4 co-assemble together in one capsid, RNA1 and RNA2 are each encapsidated separately. Quite interestingly, the total length of encapsidated genome is more or less the same in each capsid.  The BMV capsids of these three types are virtually identical, {\it i.e.}, have $T=3$ icosahedral structures consisting of 180 copies of the same protein with the same mechanical properties \cite{Zeng2017a}; see Fig.~\ref{fig:BMV_inside}. We note that the structural index T, introduced by Casper and Klug, defines the number of protein subunits in viral shells, which is 60 times the T number \cite{CASPAR1962}. Thus T = 1 and T = 3 capsids have 60 and 180 protein subunits, respectively.

\begin{figure}[t]
    \includegraphics[width=1\linewidth]{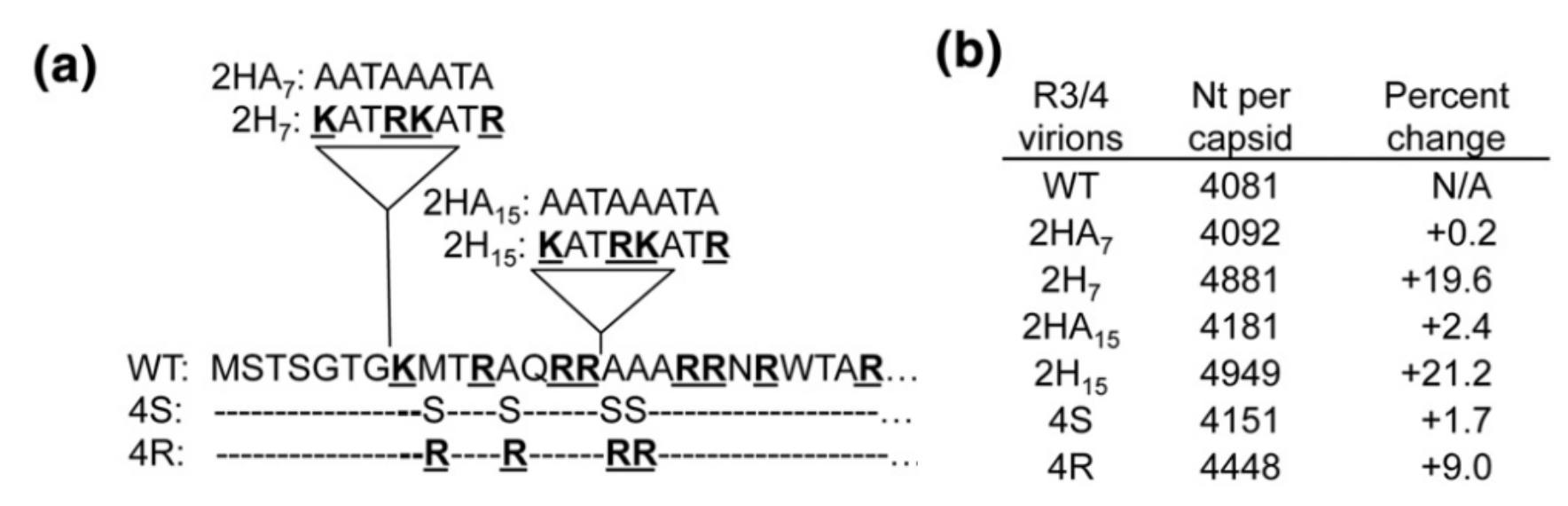}
    \caption{ (a) Schematic of the sequences of N-terminal tails of six mutants used in the experiments of Ni {\it et al.} \cite{Bogdan}. The mutants are denoted by 2HA$_7$, 2H$_7$, 2HA$_{15}$, 2H$_{15}$, 4S, and 4R. 
    The triangles denote the location of the insertions. For 2HA$_7$ and 2HA$_{15}$, eight neutral amino acids are inserted into the N-terminal. For 2H$_7$ and 2H$_{15}$, four neutral and four positive amino acids (with boldface and underlined) are inserted. The four positive amino acids are two lysines (K) and two arginines (R), leading to the increased length of N-terminal regions and also 720 additional positive charges per capsid. For 4S and 4R, the length of N-terminals remains the same. In the case of 4S four neutral amino acids (MAAA) are replaced with another four neutral amino acids and for 4R mutants, four neutral amino acids (MAAA) are replaced with four positively charged arginines (R). (b) Spectroscopic analysis of the number of nucleotides per virion.}
    \label{fig:BMV experiment}
\end{figure}

To gain more insight into the effect of electrostatic interactions, Ni {\it et al.} made several mutants to increase the number of charges on N-terminal domains. A summary of their experimental results is presented in Fig.~\ref{fig:BMV experiment}.  In one case, they inserted eight residues including four positively charged ones after residue 15 (2H$_{15}$). They also examined the impact of the length of the N-terminal without adding more positive charges but by introducing six alanines and two threonines, which are neutral (2HA$_{15}$). To examine whether the position of the insertions has an impact on the amount of packaged RNA, they repeated the aforementioned experiments but introduced insertions after residue 7 and constructed 2H$_7$ and 2HA$_7$. Furthermore, to exclusively examine the effect of the increasing charges while keeping the length of the N-terminal tail the same as the wild-type one, they replaced four uncharged residues along the tail with four arginines (4R), each containing one positive charge. They found that in all cases, the structure of capsids was almost the same even though the amount of encapsidated RNA was different. 

The spectroscopic analysis of the experiments of Ni {\it et al.} reveals that as the number of charges on the N-terminal increases, the higher amount of nucleotides per capsid is packaged \cite{Bogdan}. Nevertheless, it appears that the amount of encapsidated RNA increase does depend on other factors than the number of positive charges on the N-terminals.  While the experiments clearly indicate that electrostatics plays a major role in RNA packaging, it is not obvious whether electrostatics can explain all the effects observed in Fig~\ref{fig:BMV experiment}.  Many theoretical and experimental studies have already shown that the length of packaged RNA increases with the number of charges in N-terminal tails \cite{Li2017,Bogdan,Venky2016}, but how the amount of RNA encapsidated depends on the distribution and location of charges on the N-terminals have remained elusive.  




 In this paper we show that electrostatics is indeed able to explain at least to some extent many observed effects relevant to RNA packaging. Using the mean-field theory, we show that the charge discreteness, the location, and the distance between the charges along the N-terminal tails have a huge impact on the optimal number of nucleotides packaged.   Consistently with the experiments of Ni {\it et al.} we find that the optimal amount of packaged RNA depends on the location of charges within the peptide sequence and increases non linearly with the total number of positive charges on the capsid.

The paper is organized as follows. In the next section, we introduce the model and derive the equations that we will employ later. In Sec. III, we present our results corresponding to the non uniform charge distribution along the N-terminal tails of BMV coat proteins. Section IV discusses the impact of the length and sequence of amino acid N-terminal tails on the the length of encapsidated genomes, and  finally, we present our conclusion and summarize our findings.

\section{method}\label{sec:method}

To explore the impact of N-terminal charge distribution on the length of packaged RNA, we model RNA as a negatively charged flexible polymer. Many experiments show that RNA acts effectively as a branched polymer in solution \cite{Gopal2014, elife}. Due to the relatively weak strength of RNA base-pairing, the number of branch points of RNA can easily be modified through the interaction with the positive 
charges of virus coat proteins. Thus, we focus on the case of annealed branched polyelectrolyte, which allows the degree of branching of RNAs, a statistical quantity, to be modified \cite{Paul:13a}.   Using the mean-field theory, we calculate the free energy of the RNA confined into a spherical shell that interacts attractively with the positive charges residing on the N-terminal domains of the capsid proteins. Under the ground-state dominance approximation \cite{deGennes1979,Li_2018} where only the dominating contribution to the polymer partition function is considered, the free energy of the genome-capsid complex in a salt solution is \cite{Borukhov,Siber2008,Gonca2014,Gonca2016,Li2017}

\begin{multline} \label{free_energy}
  \beta F = \!\!\int\!\! {\mathrm{d}^3}{{{r}}}\Big[
    \tfrac{a^2}{6} |{\nabla\Psi({\bf{r}})}|^2
    +W\big[\Psi({\bf{r}}) \big]\\
    -\tfrac{\beta^2 e^2}{8 \pi \lambda_B} |{\nabla\Phi({\bf{r}})}|^2
    -2\mu\cosh\big[\beta e \Phi({\bf{r}})\big]
    + \beta \tau \Phi({\bf{r}})\Psi^2({\bf{r}})
    \Big]\\
  + \int\!\! {\mathrm{d}^2}r \Big[ \beta \rho(\bf{r}) \, \Phi({\bf{r}}) \Big].
\end{multline}
where $\beta$ is the inverse of temperature in the units of energy, $a$ is the Kuhn length of the polymer, $e$ is the elementary charge, $\mu$ is the density of monovalent salt ions, and $\tau$ is the linear charge density of chain.  The Bjerrum length $\lambda_B={e^2 \beta}/{4 \pi \epsilon}$ is about $0.7$ $nm$ for water at room temperature. The dielectric permittivity of the medium $\epsilon$ is assumed to be constant\cite{Janssen2014}. See Ref.~\cite{adsorption2015} and the Appendix of Ref.~\cite{Gonca2016} for a step by step derivation of Eq.~\eqref{free_energy}, in the absence and presence of electrostatic interactions, respectively.

The field $\Psi(\textbf{r})$ is the monomer density field and $\Phi(\textbf{r})$ is the electrostatic potential. The density of positive charges on the N-terminal tails of capsid proteins is denoted by $\rho(\textbf{r})$. The first term in Eq.~\eqref{free_energy} is the entropic cost of deviation from a uniform chain density. The last two lines of Eq.~\eqref{free_energy} are associated with the electrostatic interactions between the chain segments, the capsid, and the salt ions at the level of Poisson-Boltzmann theory \cite{Borukhov,Borukhov1,Shafir,Siber-nonspecific}.
The term $W[\Psi]$ represents the free energy density associated with the annealed branching of the polymer including the self repulsion of the polyelectrolyte \cite{Lubensky,Lee-Nguyen,Elleuch},
\begin{align} \label{W_branched}
  W[\Psi]&= -\frac{1}{\sqrt{a^3}}(f_e\Psi+\frac{a^3}{6} f_b \Psi^3)+\frac{1}{2}\upsilon \Psi^4,
\end{align}
where $f_e$ and $f_b$ are the fugacities of the end and branched points of the annealed polymer, respectively \cite{adsorption2015}, and $\upsilon$ is the effective excluded volume for each monomer. Note that the stem-loop or hair-pin configurations of RNA are counted as end points in this model. The quantity $\frac{1}{\sqrt{a^3}} f_e\Psi$ indicates the density of end points and $\frac{\sqrt{a^3}}{6} f_b \Psi^3$ the density of branch points. The expectation numbers of end and branched points, $N_e$ and $N_b$, are related to the fugacities $f_e$ and $f_b$, and can be written as
\begin{align}\label{NeNb}
N_e =- \beta f_e \frac{\partial{F}}{\partial{f_e}} \qquad {\rm and} \qquad N_b =- \beta f_b \frac{\partial{F}}{\partial{f_b}}.
\end{align}
There are two additional constraints in the system. The first one corresponds to the fact that the total number of monomers (Kuhn lengths) inside the capsid is fixed \cite{deGennes,Hone},
\begin{equation}\label{constraint}
  N = \int {\mathrm{d}^3}{\bf{r}} \; \Psi^2 ({\bf{r}}).	
\end{equation}
We impose this constraint through a Lagrange multiplier, $E$, introduced below. Second, there is a relation between the number of the end and branched points,
\begin{equation}\label{branch_constraint}
  N_e = N_b+2,
\end{equation}
as there is only a single polymer in each capsid and no closed loops within the secondary structure of an RNA are allowed. The polymer is linear if $f_b=0$, and the number of branched points increases with \textrm{increasing value of $f_b$}. For our calculations, we vary $f_b$ and find $f_e$ through Eq.~\eqref{NeNb} and Eq.~\eqref{branch_constraint}. To this end, $f_e$ is not a free parameter.

Extremizing the free energy with respect to the fields $\Psi(\textbf{r})$ and $\Phi(\textbf{r})$, subject to the constraint that the total number of monomers inside the capsid is constant (Eq.~\eqref{constraint}), we obtain three self-consistent non-linear coupled equations for the interior and exterior of the capsid,

\begin{subequations} \label{euler}
  \begin{align}
  \begin{split}
    \frac{a^2}{6} \nabla^2 \Psi (\mathbf{r})
   &=-E {\Psi} (\mathbf{r}) + \tau\beta \Phi_{in}(\mathbf{r}) \Psi(\mathbf{r})+
    \frac{1}{2} \frac{\partial W}{\partial \Psi} \label{euler_a}
    \end{split}\\
    \begin{split}
    \nabla^2 \Phi_{in} (\mathbf{r}) 
    &= \tfrac{1}{\lambda_D ^{2}}
    \sinh \big [ \Phi_{in}(\mathbf{r}) \big ]
    -  \tfrac{\tau}{{2 \lambda_D ^{2}}  \mu \beta e^2}
    {\Psi}^2 (\mathbf{r})\\
    &\hspace{8.5em}-\tfrac{1}{{2 \lambda_D ^{2}}  \mu \beta e^2} \rho(\mathbf{r})
    \label{euler_b}
   \end{split}\\
   \begin{split}
    \nabla^2 \Phi_{out} (\mathbf{r}) 
    &= \tfrac{1}{\lambda_D ^{2}}
    \sinh \big [ \Phi_{out}(\mathbf{r}) \big ] \label{euler_c}
    \end{split}
  \end{align}
\end{subequations}
where $\lambda_D=1/\sqrt{8\pi \lambda_B \mu}$ is the (dimensionless)
Debye screening length and $E$ is the Lagrange multiplier implementing the fixed monomer number inside capsid. The polymer concentration in the exterior of the
capsid is considered to be zero, $\Psi = 0$. Equations \eqref{euler} along with the constraints shown in Eqs.~\eqref{constraint} and \eqref{branch_constraint} represent a set of coupled nonlinear differential equations that, subject to appropriate boundary conditions, can only be solved numerically for the unknown parameters $f_e$ and $E$ and fields $\Psi(\textbf{r})$ and $\Phi(\textbf{r})$.

 The boundary conditions for the two coupled differential equations~\eqref{euler_b} and \eqref{euler_c} can be obtained by minimizing the free energy with respect to the $\Phi(\textbf{r})$ field on the surface of the capsid and are,
\begin{equation}
\begin{aligned}\label{eq:BC}
\hat{n}\cdot\nabla\Phi_{in}(\textbf{r})|_{r=R}&=\hat{n}\cdot\nabla\Phi_{out}(\textbf{r})|_{r=R}\\
\Phi_{in}(\textbf{r})|_{r=R}&=\Phi_{out}(\textbf{r})|_{r=R}\\
\Phi_{out}(\textbf{r})|_{r=\infty}&=0.
\end{aligned}
\end{equation}
We employ Dirichlet boundary condition $\Psi(\textbf{r})|_{r=R}=0$ for the monomer density field at the capsid wall. Because of the symmetric monomer distribution, we set $\partial_r\Psi(\textbf{r})|_{r=0}=0$. We emphasize that the derivations of all equations given in this section can be found in the Appendix of Ref.~\cite{Gonca2016}.  A more detailed derivation of the partition function and free energy for branched polymers can be found in Ref.~\cite{adsorption2015}.


\subsection{\label{sec:level2}\textbf{N-terminal tails}}
\begin{figure}[ht]\centering

\includegraphics[width=1\linewidth]{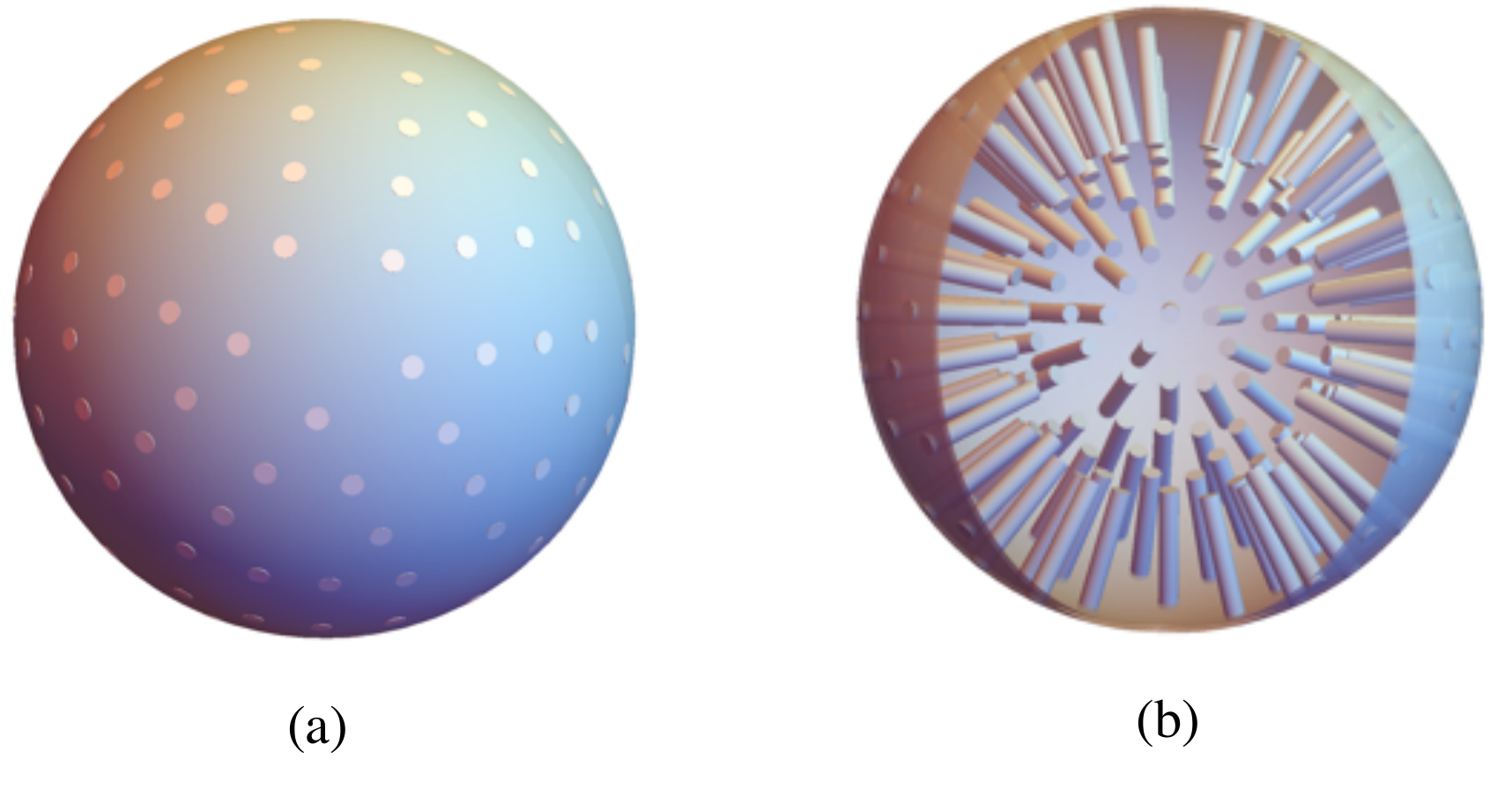}
	\caption{ (a) The white circles indicate the locations of N-terminals on a $T=3$ capsid. (b) 3D view of inside of a $T=3$ capsid with 180 protruded regions representing N-terminals. There are eight positive charges on each cylinder (N-terminal tail) in a wild-type BMV capsid. The positive charges are not shown in the figure.
	}
  \label{fig:model}
\end{figure}

 Figure~\ref{fig:BMV_inside} shows a $T=3$ structure with 180 N-terminal tails extending into the interior of the capsid, distributed with icosahedral symmetry. Because of the repulsion between the positive charges residing on the N-terminal tails, and the fact that RNA wraps around them, we assume that the N-terminal tails take an extended configuration. To this end, we model the N-terminal tails of BMV capsids as solid cylinders; see Fig.~\ref{fig:model}(b). We note that the charged tails are placed inside the capsid, and we will use the same boundary conditions for them as those given in Eq.~\eqref{eq:BC} at the surface.

In the next section we will examine the impact of different charge distributions along N-terminal domains on the optimal genome length, which we will compare with the experimental results presented in Fig.~\ref{fig:BMV experiment}. Since most of the positive charges are residing on the N-terminal tails, we consider that the charges of the coat proteins are only distributed in the cylindrical regions with no charges on the capsid wall.

For simplicity, we first consider a $T=1$ capsid with only two positive charges on each of its 60 N-terminal tails and then focus on the $T=3$ capsid of BMV.  
\section{\label{sec:level1-2}Results}

\subsection{\label{sec:level2-1}\textbf{A capsid with 60 tails ($T=1$)}}

To obtain the optimal length of encapsidated genome in a $T=1$ shell, we numerically solve the nonlinear coupled differential equations \eqref{euler_a}, \eqref{euler_b}, and \eqref{euler_c}, subject to the constraints given in Eqs.~\eqref{constraint} and \eqref{branch_constraint}. We operate on the nonlinear coupled differential equations with the finite element method and deal with the convergence issue employing the Newton method \cite{BangerthHartmannKanschat2007, bathe1996finite, NoceWrig06}. 

\begin{figure}[ht]
\includegraphics[width=1\linewidth]{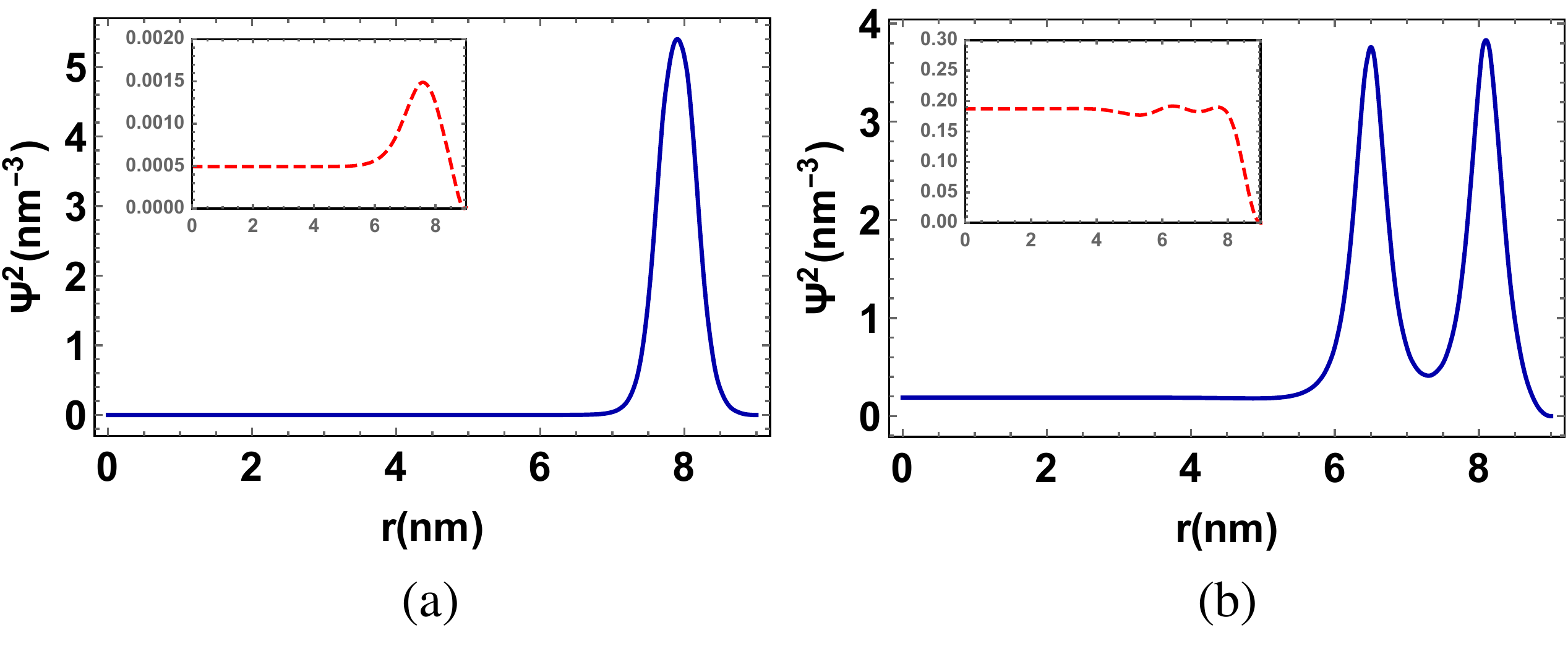}
    \caption{ Genome density profile inside a $T = 1$ capsid as a function of the distance from the capsid center. The solid lines in the figure show the profiles along N-terminal tails, but the dashed graphs correspond to the direction without N-terminal tails (inset). (a) The plot illustrates the profile when the distance between the two charges is  $0.2$ $nm$ with the total number of monomers $N=658$. (b) The plot corresponds to the profile when the distance between the two charges is $1.4$ $nm$ with $N=680$. See Fig.~\ref{fig:distance}(a) for a schematic view of charge distributions.  The length of the tail is $4$ $nm$, and the size of each charged region is $0.2$ $nm$.  The polymers are branched with $f_b = 3.86$. The other parameters are salt concentration $\mu = 500$ $mM$, the capsid radius $R = 9$ $nm$, and the total charge on N-terminals $Qc = 120$.
    }
    \label{fig:T=1genomeprofile}
\end{figure}

After finding the solutions for the fields $\Psi(\textbf{r})$ and $\Phi(\textbf{r})$ we insert them into Eq.~\eqref{free_energy} to obtain the free energy of the polymer-capsid complex, $F$ \cite{Gonca2014,Gonca2016,GoncaPRL2017}. To obtain the encapsidation free energy, we need to calculate the free energy of a polymer free in solution and that of a positively charged shell and then subtract them both from the polymer-capsid complex free energy, $F$, given in Eq.~\eqref{free_energy}.  The capsid self-energy [F (N = 0)] due to the electrostatic interactions is calculated through Eqs.~\eqref{euler} and \eqref{eq:BC} in the limit as N → 0 and should be explicitly subtracted from the polymer-capsid complex free energy, $F$. We emphasize that the focus here is on the solution conditions in which the capsid proteins can self-assemble in the absence of the genome. We also note that previous works have shown that the free energy associated with a free chain (both linear and branched) is negligible under most experimental conditions \cite{Gonca2016, Siber2008}.

\begin{figure}[t]\centering

\includegraphics[width=1\linewidth]{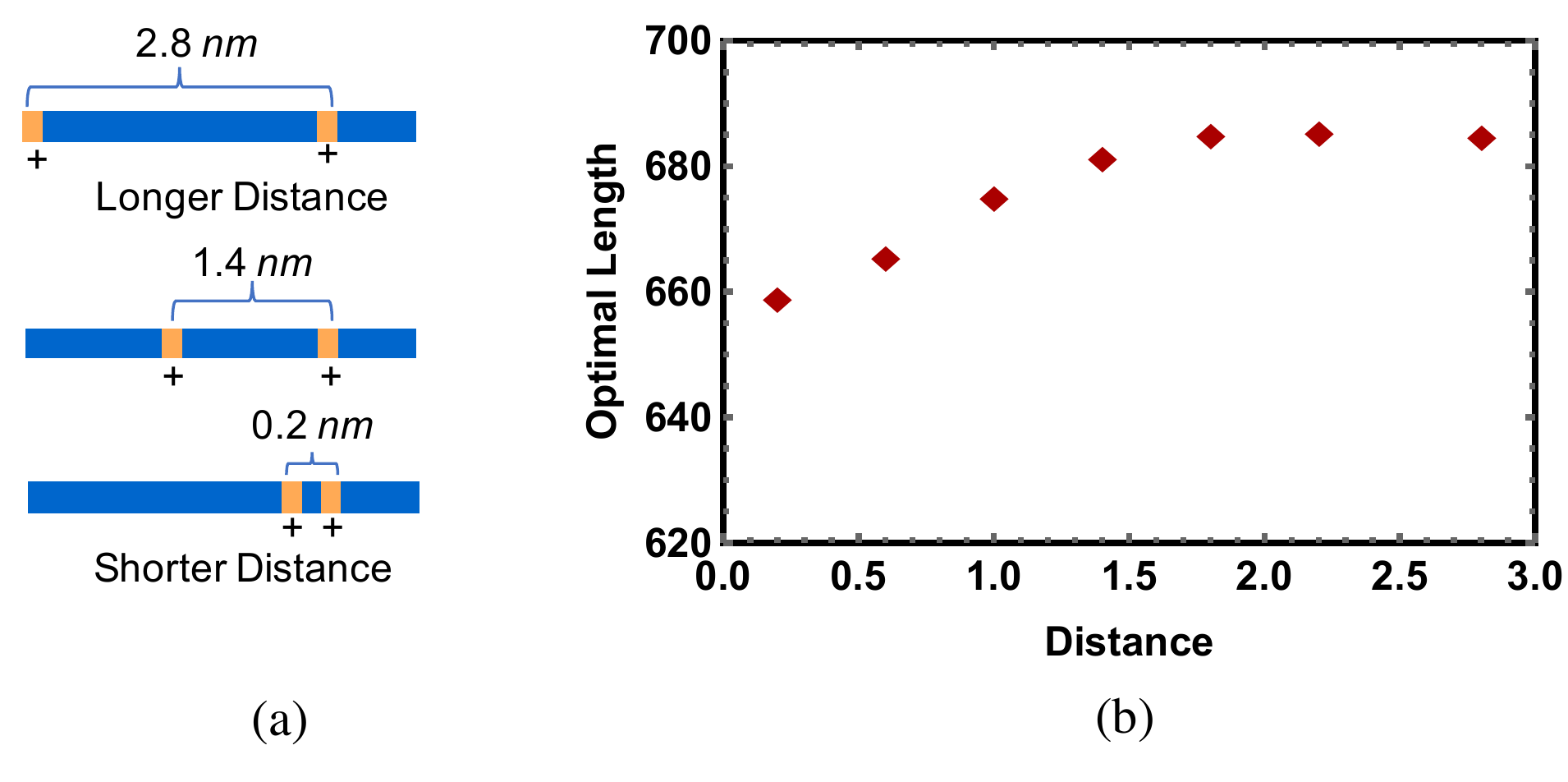}
  \caption{(a) Schematic of an N-terminal tail.  The distance between two positive charges along the N-terminal domain increases from bottom to top. Each yellow rectangle is $0.2$ $nm$ and denotes
  one positive charged amino acid.  The smallest distance between the two charges is $0.2$ $nm$. From the shortest to the longest distance, we examine seven different cases. The largest distance between the two charges is $2.8$ $nm$.  The charge on the right side is next to the wall and its position is fixed.  (b) Optimal length of RNA encapsulated as a function of the distance between two charges for a capsid with radius $R = 9$ $nm$, the tail length $4$ $nm$, and salt concentration $\mu = 500$ $mM$. RNA is modeled as an annealed branched polymer. 
  }
  \label{fig:distance}

\end{figure}

The results of our numerical calculations are given in Fig.~\ref{fig:T=1genomeprofile} as a plot of the polymer concentration profile vs~$r$, the distance from the center of the shell for a branched polymer with the radius of capsid $R = 9$ $nm$ at $\mu = 500$ $mM$ salt concentrations. The total number of charges in the capsid is $Qc = 120$ with two charges on each N-terminal tail. The length of N-terminal is $4$ $nm$ and the size of each charge is $0.2$ $nm$ (see Fig.~\ref{fig:distance}(a)). 

Figure~\ref{fig:T=1genomeprofile}(a) shows the genome profile if the distance between the two positive charges along the N-terminal tails is $0.2$ $nm$ while Fig.~\ref{fig:T=1genomeprofile}(b) corresponds to when the distance between the charges is $1.4$ $nm$; see Fig.~\ref{fig:distance}(a) for a schematic presentation of the distribution of charges in both cases. Note that the charged amino acids are yellow and neutral ones are blue in Fig.~\ref{fig:distance}(a).   The optimal number of monomers enclosed in the shell for Fig.~\ref{fig:T=1genomeprofile}(a) is $N=658$ and for Fig.~\ref{fig:T=1genomeprofile}(b) is $N=680$.  The figure clearly shows that the polymer concentration is higher at the positions where the positive charges are located along the tails. When the distance between two charges is less than the Debye length $\lambda_D=0.438$ $nm$, there is only one maximum in the profile. As the distance between the charges increases and goes beyond two Debye lengths, the genome density profile between the two charges goes almost to zero.

It is important to note that we have previously studied the impact of the number of branched points, which is closely connected to the $f_b$ value, on the length of the encapsidated genome and found that the length of the genome increases with $f_b$ \cite{Gonca2014}. Since our focus in this paper is only on the effect of charge distribution along the N-terminals, we set $f_b=3.86$ for all the calculations presented here. In a previous paper, we found that this value of $f_b$ would create a similar number of branch points to the case in the wild-type BMV genome \cite{Gonca2014}. The value of $f_b$ does not play an important role in our findings of the effect of N-terminal charge distribution.

\begin{figure}[ht]\centering
  \includegraphics[width=0.85\linewidth] {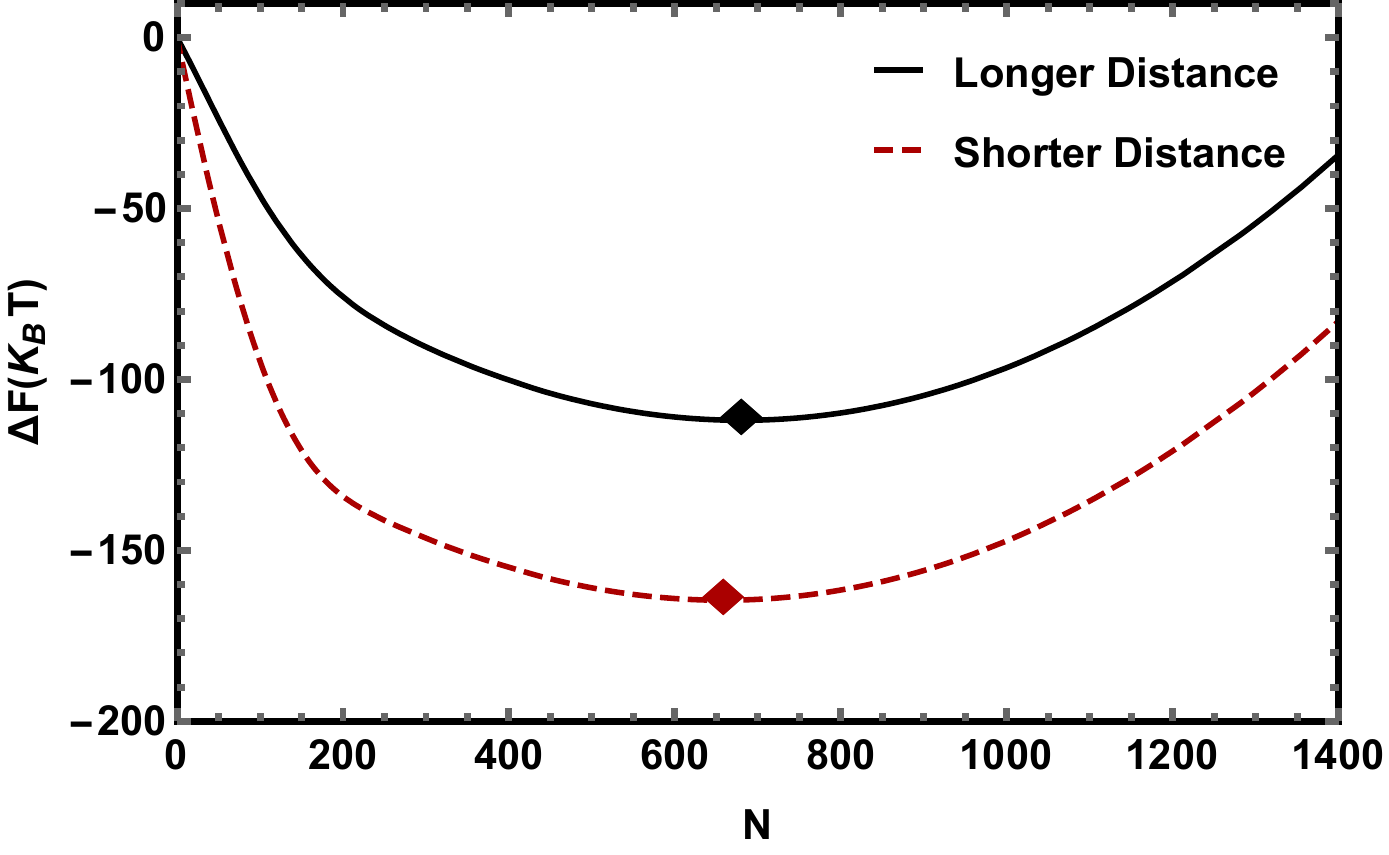}
   
  \caption{Encapsulation free energy as a function of monomer number, $N$. The dashed line corresponds to the case in which the distance between the two charges is short ($d=0.2$ $nm$) and the solid curve to when the distance between the charges is a little bit longer ($d=1.4$ $nm$); see Fig.~\ref{fig:distance}(a) for a schematic of two charge distributions. The other parameters are the capsid radius $R = 9$ $nm$, the tail length $4$ $nm$, and salt concentration $\mu = 500$ $mM$ and the total positive charge on the capsid is $Qc = 120$. RNA is modeled as an annealed branched polymer and its fugacity is $f_b = 3.86$. The optimal number of packaged monomers for $d=0.2$ $nm$ it is 658 while for $d=1.4$ $nm$ it is 680. 
 }\label{fig:free energy distanceT1}

\end{figure}

Figure~\ref{fig:free energy distanceT1} shows the encapsulation free energy as a function of $N$, the number of monomers, for a $T = 1$ structure. The dashed line in the figure corresponds to the case in which the distance between the charges is $0.2$ $nm$ and solid lines to when the distance between the charges is $1.4$ $nm$; see Fig.~\ref{fig:distance}(a) for a schematic of two charge distributions. As illustrated in Fig.~\ref{fig:free energy distanceT1}, when the charges are closer to each other, the free energy of the system is lower; however, the minimum of the free energy moves toward longer chains as the distance between the charges increases.

Figure~\ref{fig:distance}(b) shows the optimal length of encapsidated RNA as a function of the distance between two charges along the N-terminal domains. One charge is placed at the end of the N-terminal tail next to the capsid wall, but the location of the other varies from the wall all the way to the tip. The figure clearly shows that as the distance between charges increases, the optimal length of the genome increases too. Thus, the location of charges along the N-terminal domains has an impact on the amount of the polymer packaged. It appears as the distance between the charges goes up, at some point the optimal length of the packaged genome saturates and does not keep increasing. 
A careful examination of the first term in Eq.~\eqref{free_energy} shows that for this size of capsid and charge distribution, the optimal genome density is too small and the impact of entropy is not strong enough to have a significant role in the optimal length of the genome. 
 As the distance between the charges increases and becomes more than two Debye lengths ($\lambda_D = 0.438$ $nm$ for $\mu=500$ $mM$), the electrostatic interaction becomes very weak between the two charges. Thus, the genome will be mostly adsorbed in the close proximity of each positive charge along the peptide. Note that even though entropy prefers a uniform genome density, the electrostatic interaction is much stronger and thus the optimal length of encapsidated genome first increases with the distance between the charges and then it remains more or less constant.

\subsection{\label{sec:level2-2}\textbf{A capsid with 180 tails ($T=3$)}}

\begin{figure}[ht]
    \includegraphics[width=\linewidth]{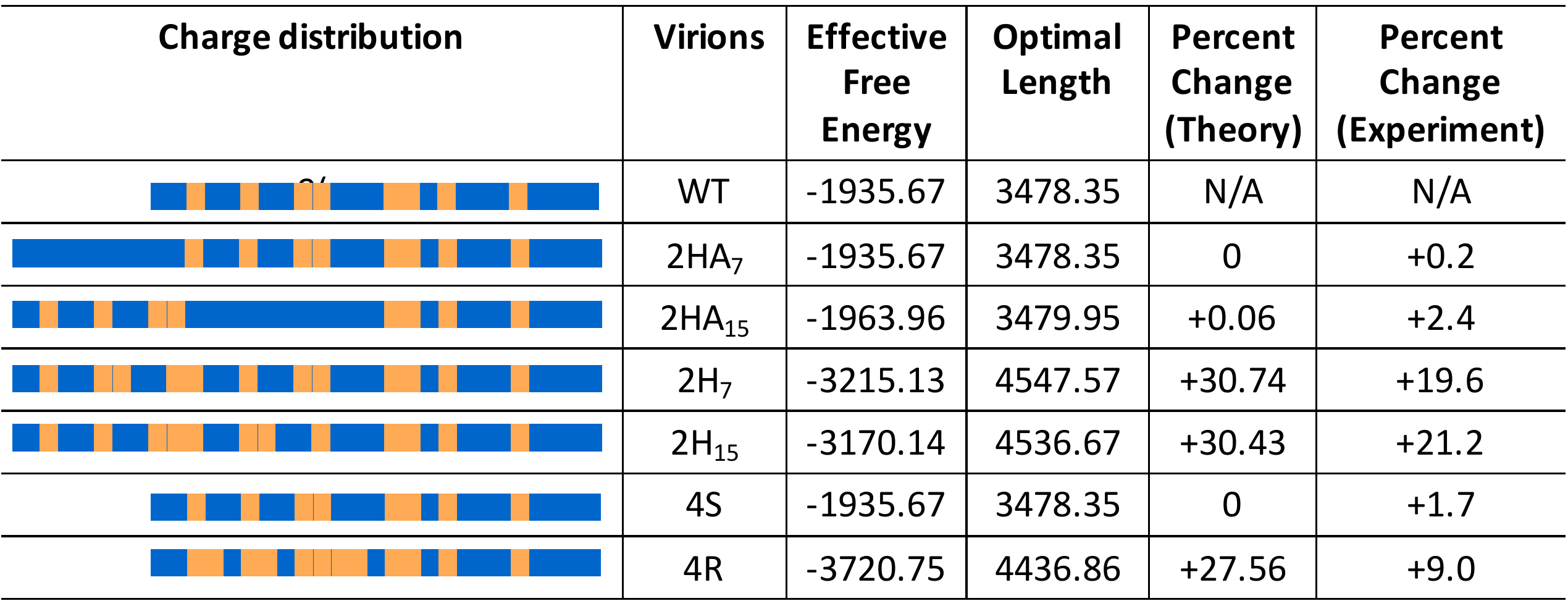}
    \caption{Table of seven charge distributions along N-terminals where each yellow rectangle represents a positively charged amino acid and blue triangles neutral ones. The table includes the optimal encapsulation free energy of the RNA confined into a spherical shell, the optimal length of encapsulated RNA, percent change (theory) of optimal length compared to the wild-type BMV and the percent change (experiment) from Fig.~\ref{fig:BMV experiment}. The salt concentration is $500$ $mM$. The radius of the capsid is $12$ $nm$. For wild type, 2HA$_7$, 2HA$_{15}$, and 4S, the total charge on capsid is $Qc = 1440$ but for 2H$_7$, 2H$_{15}$, and 4R it is $Qc = 2160$. The tail length for wild type, 4S, and 4R is $5$ $nm$ while for 2HA$_7$, 2HA$_{15}$, 2H$_7$, and 2H$_{15}$ it is $6.5$ $nm$. 
    The Debye length $\lambda_D$ for $500$ $mM$ is $0.438$ $nm$.}
    \label{fig:BMV simulation}
\end{figure}

We now examine the impact of charge distribution along the N-terminal domain for a $T=3$ capsid with 180 N-terminal tails.  More specifically, we focus on the self-assembly studies of Ni {\it et al.} in which the impact on the length of packaged RNA of the location and distribution of positive charges along the N-terminal domains of BMV capsid proteins were studied \cite{Bogdan}. Figures~\ref{fig:BMV experiment}(a) and~\ref{fig:BMV experiment}(b) show the distribution of charges along N-terminal domains and the length of encapsulated RNA for different mutants, respectively. The schematic of the charge distribution along the N-terminals for various mutants and wild-type capsid proteins based on our model is illustrated in the left column of Fig.~\ref{fig:BMV simulation}.  The length of N-terminal is set equal to $5$ $nm$ for the wild type and $6.5$ $nm$ for the mutants with eight extra amino acids. We assume all amino acids have the same size, which is set equal to $0.2$ $nm$. The charged amino acids are yellow and neutral ones are blue as before.
 
  \begin{figure}[ht]\centering

\includegraphics[width=1\linewidth]{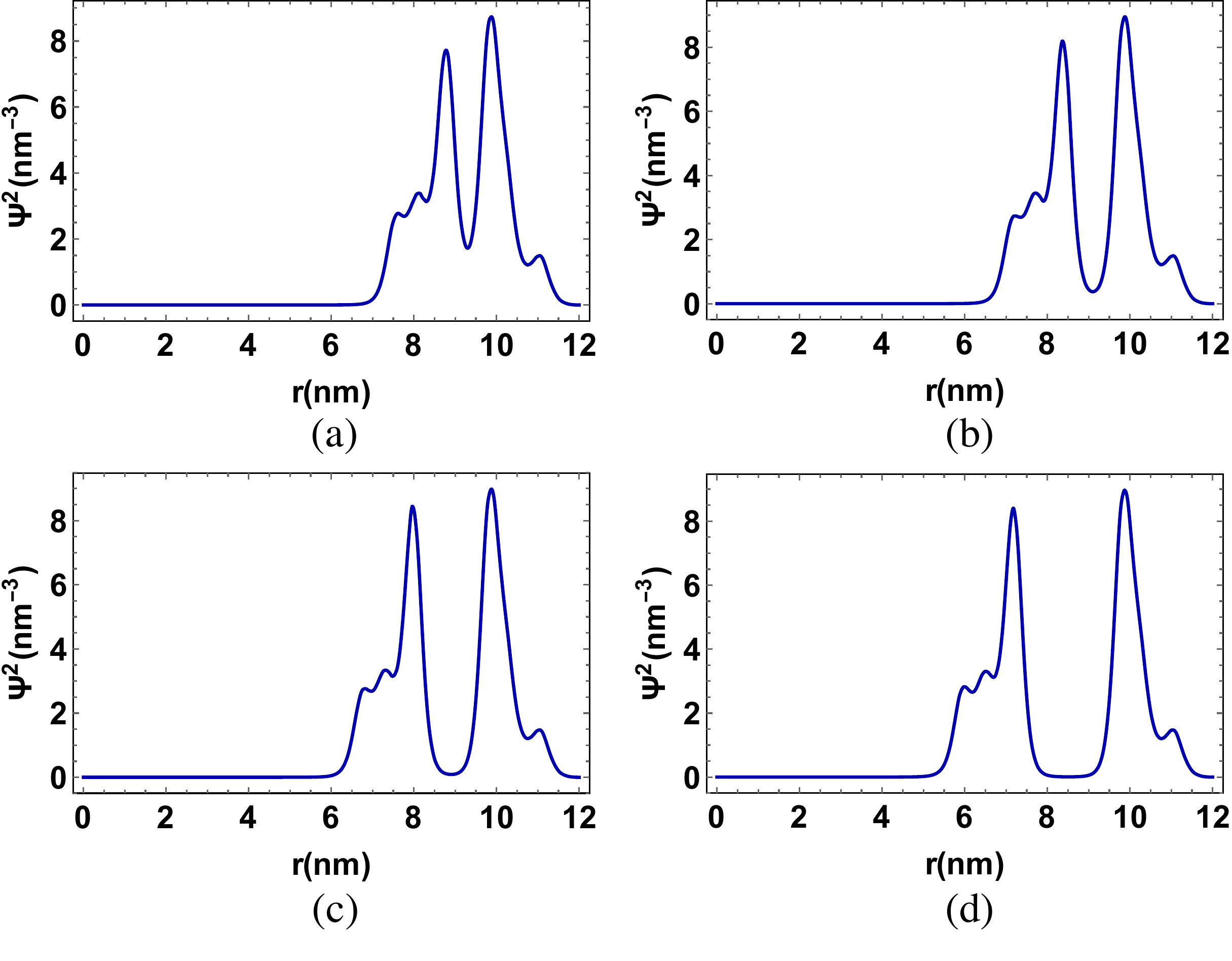}
  \caption{The genome density profile vs r, the distance from the center of the capsid for four different charge distributions along the N-terminal domains: (a) wild type (WT), (b) 2HA$_{15}$(M1), (c) 2HA$_{15}$(M2), and (d) 2HA$_{15}$. The first column in Figs.~\ref{fig:BMV simulation} and \ref{fig:BMV simulation extra(2)} shows the schematics of N-terminal tails for each case. The peaks in the RNA profiles correspond to the position of positive charges along the N-terminal tails. As the distance between the charges located in the middle of N-terminal tails increases, the density of genome between the two peaks goes lower. However, the amount of RNA between two peaks due to the entropic contribution and the range of electrostatic interaction do not drop to zero in the case of 2HA$_{15}$(M1) (b).  The genome density between the two peaks becomes smaller for 2HA$_{15}$(M2) (c) and becomes almost zero for 2HA$_{15}$.}
  \label{fig:BMV genome profile}

\end{figure}
 
\begin{figure}[ht]
    \includegraphics[width=\linewidth]{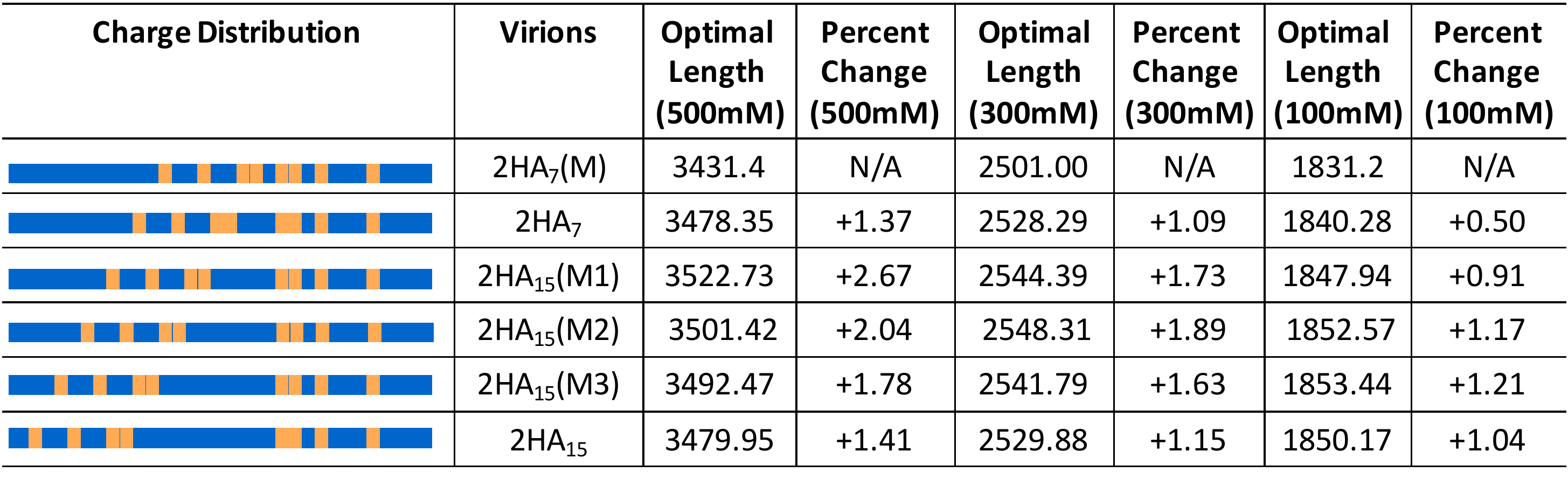}
    \caption{Table of six different charge distributions along N-terminals. As before each yellow rectangle represents an amino acid with a positive charge and blue rectangles represent neutral amino acids. The table includes the optimal length of encapsulated RNA for three different salt concentrations, $500$ $mM$, $300$ $mM$, and $100$ $mM$.  The distance between the fourth positive charge (the fourth yellow rectangle) and the fifth positive charge from top to bottom is $0.2$ $nm$, $0.6$ $nm$, $1.0$ $nm$, $1.4$ $nm$, $1.8$ $nm$, and $2.2$ $nm$.
    The percent change (theory) of the optimal length of encapsidated RNA for each mutant relative to the RNA encapsidated by mutant 2HA$_{7}$(M) is also presented in the table.  The capsid radius is $12$ $nm$ and the tail length is $6.5$ $nm$ with total charges on the capsid $Qc = 1440$. Debye length is $\lambda_D=0.979$ $nm$ for $\mu=100$ $mM$, $\lambda_D=0.565$ $nm$ for $\mu=300$ $mM$ and $\lambda_D=0.438$ $nm$ for $500$ $mM$.}
    \label{fig:BMV simulation extra(2)}
\end{figure}

Following the same procedures as described above for a $T=1$ structure, we first obtain the genome profile for a given number of nucleotides and then use it to calculate the free energy of the system. Figure \ref{fig:BMV genome profile} shows the genome profiles for the wild type, 2HA$_{15}$, and two other mutant proteins. The schematic of charge distribution for each case is illustrated in Figs.~\ref{fig:BMV simulation} and \ref{fig:BMV simulation extra(2)}.  The total number of monomers in each plot in Fig.~\ref{fig:BMV genome profile} is $N=1390$, and the total number of charges in all capsids is $Qc = 1440$. There are eight positive charges on each N-terminal tail, whose length is $6.5$ $nm$ long for mutants and $5$ $nm$ for wild-type proteins. The genome is considered to be a branched polymer ($f_b=3.86$).

\begin{figure}[ht]
    \includegraphics[width=\linewidth]{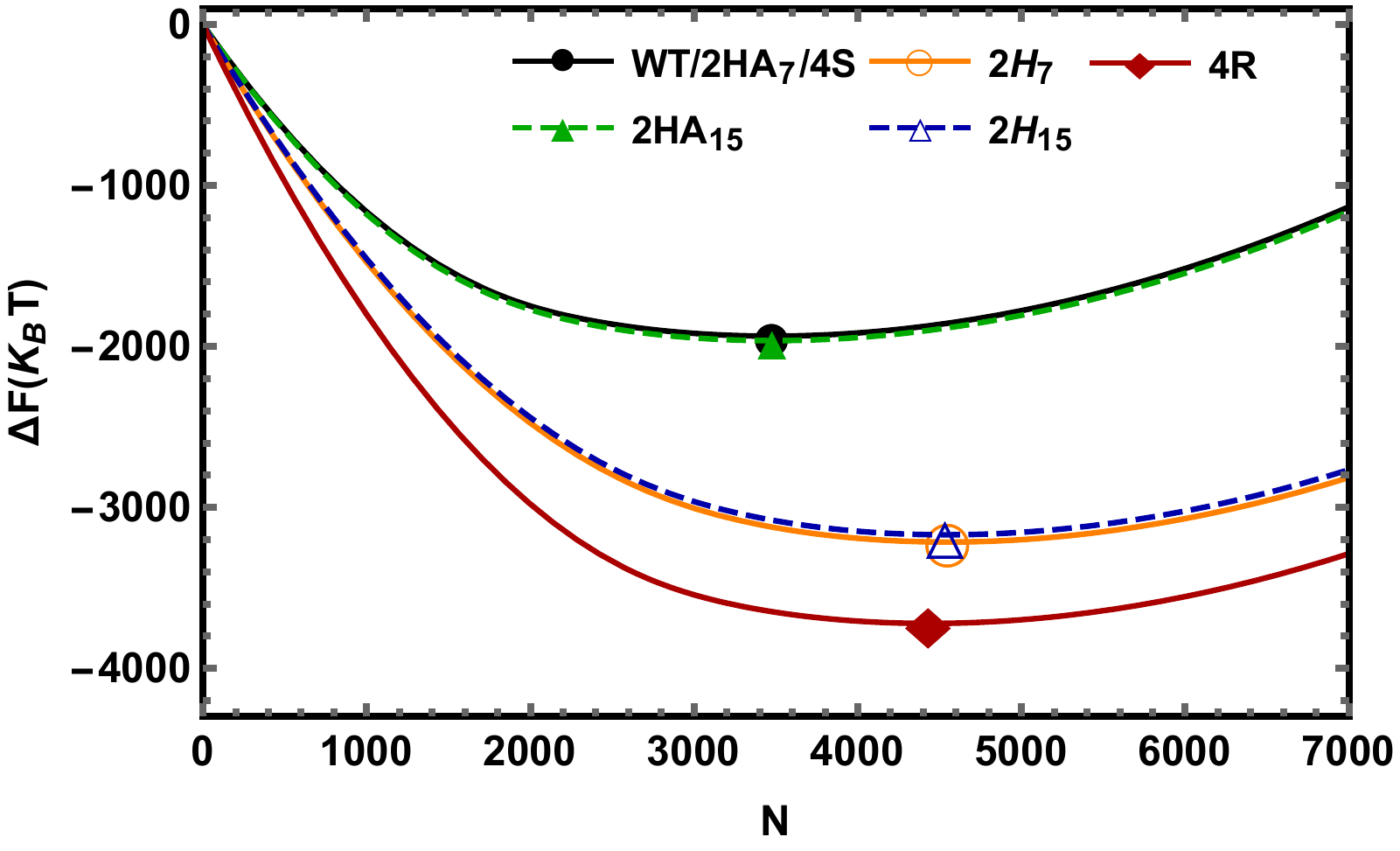}
    \caption{The encapsidation free energy as a function of monomer numbers for the mutants presented in Fig.~\ref{fig:BMV simulation}. 2HA$_7$ and 4S have the same free energy as wild type. The additional length inserted in 2HA$_7$ does not have a huge impact on the optimal encapsidated genome because it does not modify the distance between the charges along the N-terminal tails; see Fig.~\ref{fig:BMV simulation}. The capsid radius is $12$ $nm$ and the tail length is $6.5$ $nm$ with total charges on the capsid $Qc = 1440$. The salt concentration is $500$ $mM$.
    }
    \label{fig:BMV simulation free energy}
\end{figure}

Figure~\ref{fig:BMV simulation free energy} shows the free energy of a branched polymer packaged by the wild-type and mutant proteins of Fig.~\ref{fig:BMV simulation}. The symbols in the figure correspond to the optimal genome length for each case. The figure reveals that the encapsulation free energy of the wild-type, 2HA$_7$, 4S, and  2HA$_{15}$ are almost the same. Note that all these mutants have the same number of charges on their capsids. The values of the minimum free energy, the corresponding optimal genome length, and the percent change (theory and experiment) of encapsulated genome compared to the wild-type case are presented in Fig.~\ref{fig:BMV simulation}.

Consistent with the experimental data presented in Fig.~\ref{fig:BMV experiment} and the last column of Fig.~\ref{fig:BMV simulation}, our theoretical calculations show that as the number of positive charges on the N-terminal tails increases, the optimal length of the genome increases too.  The mutants 4R, 2H$_7$, and 2H$_{15}$ have four extra positive charges compared to wild-type proteins and they all encapsidate longer genomes. Both mutants 2H$_7$ and 2H$_{15}$ have longer tails compared to 4R, and our results show that they encapsidate longer genomes, consistent with the experimental findings.  Thus the length of N-terminal tails influences the amount of packaged RNA. 

While there are many similarities between the experiments presented in Fig.~\ref{fig:BMV experiment} and our theoretical results shown in Fig.~\ref{fig:BMV simulation}, there are also some differences. The comparison of the experiment and theory reveals that more genome is encapsidated by 2HA$_{15}$ proteins compared to wild-type or 2HA$_7$ proteins, which is not observed in our calculations. Note that to perform the numerical calculations, we consider that all amino acids have the same effective size ($0.2$ $nm$), and the Debye length in our system is $\lambda_D=0.438$ $nm$.  Since the parameter landscape is quite vast and there are several unknowns, instead of changing the size of each amino acid, we modify the distance between the fourth and fifth charged amino acids in the N-terminal tail of the mutant 2HA$_{15}$. More specifically, we systematically increase the distance between the fourth and fifth positive charges from $0.2$ $nm$ to $2.8$ $nm$  where the 8 amino acids were inserted for the case of the mutant 2HA$_{15}$ and then calculate the optimal length of encapsidated genome for three different salt concentrations of $\mu=100$, $300$ and $500$ $mM$.  As illustrated in Fig.~\ref{fig:BMV simulation extra(2)}, the optimal length of encapsidated genome depends on both the distance between the fourth and fifth positively charged amino acids and the salt concentration. The figure reveals that as the distance increases from $0.2$ to $2.2$ $nm$, the optimal length of the encapsidated genome first increases and then later decreases. 

\begin{figure}[ht]
  \includegraphics[width=\linewidth]{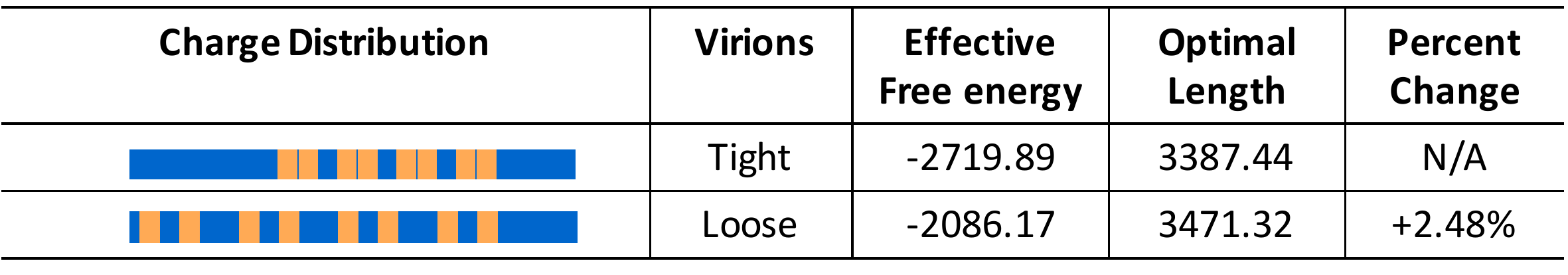}
   \caption{Schematic of two N-terminal tails with different charge distributions. As before each yellow rectangle represents an amino acid with a positive charge and effective size of $d=0.2$ $nm$ and blue rectangles represent neutral amino acids but with the same size. The table includes the effective encapsulation free energy of the RNA confined into a spherical shell, the optimal length of encapsulated RNA, and the percent change (theory) of optimal length of packaged RNA with respect to the first charge distribution. The salt concentration is $\mu=500$ $mM$, and tail length is $4.5$ $nm$ for both cases. The total charge on the capsid is $Qc=1440$. When the charges are distributed more evenly, the optimal length of the encapsidated genome increases. For the first line of the table, the distance between yellow rectangles is either zero or $0.2$ $nm$, while for the second one it is either $0.2$ $nm$ or $0.4$ $nm$.
    }
    \label{fig:BMV simulation extra extra}
\end{figure}

To gain more insights into the experimental results, we also examined the impact on the optimal polymer length of a uniform charge distribution along the N-terminals versus a tight one as presented in Fig.~\ref{fig:BMV simulation extra extra}.   As shown in the figure, for a given tail length and number of positive charges, when the charges are distributed more uniformly along the N-terminals, the optimal length of encapsidated genome becomes longer.

\section{\label{sec:level1-3}Discussion and Summary}

Despite the fact that many experiments have shown that the number of nucleotides packaged by capsid proteins increases with the number of charges on N-terminal tails, how the amount of encapsidated RNA depends on the distribution of the charges along and the length of the N-terminal domain of capsid proteins is not well understood.   Our results presented in Fig.~\ref{fig:distance}(b) for T=1 capsid and Fig.~\ref{fig:BMV simulation} and Fig.~\ref{fig:BMV simulation extra(2)} for T=3 viruses show that the electrostatic interaction alone is not sufficient to explain the dependence of the amount of packaged RNA on the amino sequence of N-terminal tails in the BMV experiments \cite{Bogdan}.  For example, the amount of packaged RNA is different for the mutant 2HA$_{15}$ and 2HA$_7$ as illustrated in Fig.~\ref{fig:BMV experiment} whereas both have the same number of charges, very similar charge distribution, and the same peptide length. This reveals the importance of specific interactions that depend on the exact type of amino acids, RNA secondary or tertiary structures, and packaging sequences or signals, which involve the highly specific, nonelectrostatic interactions between sections of RNA and capsid proteins \cite{Haganpackaging,Stockley2013,Patel2014}.  Our mean-field theory does not include this effect and thus cannot explain the experimental observation due to specific interactions; nevertheless, our theory can describe how the length of N-terminal tails and distribution of charges along the peptide control the amount of RNA packaged by BMV capsid proteins, consistent with the experimental data.

The simple case of two charges on the N-terminal tails of the $T=1$ capsid (Fig.~\ref{fig:distance}) shows clearly that when the distance between two positive charges increases, the optimal length of RNA encapsulated into the capsid also increases. A careful examination of Eq.~\eqref{free_energy} shows that the length of the encapsulated polyelectrolyte increases with the distance in order for the chain to be uniformly distributed between the two charges, lowering the entropy contribution (the first term in Eq.~\eqref{free_energy}) as much as possible. Figure~\ref{fig:distance} shows that the optimal length of the genome saturates and remains more or less constant beyond a certain distance between the two charges. This is mainly due to the fact that the optimal length of the genome for $T=1$ is such that the density of the genome is low. When the distance between the charges is more than two Debye lengths ($\lambda_D = 0.438$ $nm$ for $\mu=500$ $mM$), the electrostatic interaction becomes very weak between the distant charges. It appears that the chain then prefers to reside only in the immediate vicinity of each positive charge along the peptide. More specifically, as the distance between the charges increases, the electrostatic does not promote encapsidation of longer genomes. 
Thus, the optimal length of the genome first increases and then it remains constant even if the distance between the charges increases further.

Figure~\ref{fig:BMV simulation} shows that the case for the $T=3$ structure relevant to BMV experiments is more complex. As seen in the figure, the mutant 4R whose charge is increased by substitution instead of insertion (keeping the length constant) has less encapsidated RNAs than do 2H$_7$ and 2H$_{15}$, while all three mutants have the same number of charges on their tails. Our calculations reveal that since 2H$_7$ and 2H$_{15}$ have longer N-terminal tails, a longer genome is necessary for the chain to {\it uniformly} wrap around the tail keeping the entropic contribution in Eq.~\eqref{free_energy} low.  However, the difference between the length of the genome encapsidated by wild-type proteins and the mutant 2HA$_{15}$ proteins whose N-terminal length is increased by insertion of eight neutral amino acids is not as pronounced in our theory as in the experiments.  This could be explained at least in part by the distance between the charged amino acids in the peptide. To understand the impact of the distance between the charges, we systematically examined the effect of the distance between the charges in the middle of the N-terminal tail as illustrated in Fig.~\ref{fig:BMV simulation extra(2)}.  The results presented in the figure are quite intriguing as the optimal length of the genome first increases and then decreases for the three different salt concentrations presented in the figure.  

The large distance between the two charges along the N-terminal tail provides more space for the genome to reside. The careful examination of Eq.~\eqref{free_energy} shows that due to the entropic consideration, the genome will be distributed more or less uniformly along the N-terminal leading to the packaging of longer genomes. However, as the distance between the charges increases and goes beyond two Debye length ($\lambda_D=0.438$ $nm$ for $\mu=500$ $mM$, $\lambda_D=0.565$ $nm$ for $\mu=300$ $mM$ and $\lambda_D=0.979$ $nm$ for $\mu=100$ $mM$), the optimal length of RNA becomes shorter.  This effect can be well understood by investigating the genome profiles presented in Fig.~\ref{fig:BMV genome profile}.   When the distance between the fourth and fifth charges is very large, there will be two distinct peaks in the genome profile with almost no nucleotides between the charges indicating that the negatively charged RNA prefers to be localized mainly around the positive charges.
Figure \ref{fig:BMV simulation extra(2)} indicates that as the distance between the charges increases, at some point the optimal length of encapsidated RNA decreases resulting in the lower polymer density, which also reduces the entropy cost of formation of two completely separate peaks.  
Since the Debye length is longer for lower salt concentrations, the optimal length of the genome starts decreasing at $d=1.8$ $nm$ for $\mu=100$ $mM$ (2HA$_{15}$(M3)), $d=1.4$ $nm$ for $\mu=300$ $mM$ (2HA$_{15}$(M2)) and $d=1.0$ $nm$ for $\mu=500$ $mM$ (2HA$_{15}$(M1)). While the behavior is the same for all three salt concentrations, the effect is less pronounced as the salt concentration decreases.  Figure~\ref{fig:BMV simulation extra extra} further supports that for a given length and number of positive charges, the more uniformly charges are dispersed along the N-terminals, the longer the optimal length of the encapsidated genome becomes.

We emphasize that the goal of this paper has been to qualitatively explain the experimental results and to explore the impact of entropy and electrostatic interaction that depend on the distance between the charges and not the details of protein structures. A better quantitative comparison between the experiments and theory can be obtained if the theory includes many other effects such as counter-ion condensation, the presence of divalent ions, the structure of proteins, and the packaging signals discussed above.   

In summary, in this paper we explore whether the variation in RNA packaging by BMV mutants observed in the experiments of Ni {\it et al.} and presented in Fig.~\ref{fig:BMV experiment}~\cite{Bogdan} can be understood by the mean-field theory incorporating electrostatics, excluded volume interaction and RNA conformational entropy. In particular, we have calculated, as a function of the number and location of charges in the peptide tails, the free energy of an RNA confined in a spherical shell interacting with the N-terminal tails and ions.  We find that the combined effect of the electrostatic interaction and the genome entropy considerations can shed light on many experimental data relevant to BMV assembly.  While our mean-field theory cannot explain all the experimental data, we have been able to show that the location and the distance between charges along the N-terminal tails significantly influence the amount of packaged RNA. Understanding the factors contributing to the virus assembly and RNA packaging will pave the path for interfering with the different stages of the virus life cycle.

\section*{Acknowledgments}
This work was supported by the National Science Foundation through Grant No. DMR-1719550.


\begin{thebibliography}{42}%
\makeatletter
\providecommand \@ifxundefined [1]{%
 \@ifx{#1\undefined}
}%
\providecommand \@ifnum [1]{%
 \ifnum #1\expandafter \@firstoftwo
 \else \expandafter \@secondoftwo
 \fi
}%
\providecommand \@ifx [1]{%
 \ifx #1\expandafter \@firstoftwo
 \else \expandafter \@secondoftwo
 \fi
}%
\providecommand \natexlab [1]{#1}%
\providecommand \enquote  [1]{``#1''}%
\providecommand \bibnamefont  [1]{#1}%
\providecommand \bibfnamefont [1]{#1}%
\providecommand \citenamefont [1]{#1}%
\providecommand \href@noop [0]{\@secondoftwo}%
\providecommand \href [0]{\begingroup \@sanitize@url \@href}%
\providecommand \@href[1]{\@@startlink{#1}\@@href}%
\providecommand \@@href[1]{\endgroup#1\@@endlink}%
\providecommand \@sanitize@url [0]{\catcode `\\12\catcode `\$12\catcode
  `\&12\catcode `\#12\catcode `\^12\catcode `\_12\catcode `\%12\relax}%
\providecommand \@@startlink[1]{}%
\providecommand \@@endlink[0]{}%
\providecommand \url  [0]{\begingroup\@sanitize@url \@url }%
\providecommand \@url [1]{\endgroup\@href {#1}{\urlprefix }}%
\providecommand \urlprefix  [0]{URL }%
\providecommand \Eprint [0]{\href }%
\providecommand \doibase [0]{http://dx.doi.org/}%
\providecommand \selectlanguage [0]{\@gobble}%
\providecommand \bibinfo  [0]{\@secondoftwo}%
\providecommand \bibfield  [0]{\@secondoftwo}%
\providecommand \translation [1]{[#1]}%
\providecommand \BibitemOpen [0]{}%
\providecommand \bibitemStop [0]{}%
\providecommand \bibitemNoStop [0]{.\EOS\space}%
\providecommand \EOS [0]{\spacefactor3000\relax}%
\providecommand \BibitemShut  [1]{\csname bibitem#1\endcsname}%
\let\auto@bib@innerbib\@empty
\bibitem [{\citenamefont {Bancroft}(1970)}]{Bancroft}%
  \BibitemOpen
  \bibfield  {author} {\bibinfo {author} {\bibfnamefont {J.~B.}\ \bibnamefont
  {Bancroft}},\ }\href@noop {} {\bibfield  {journal} {\bibinfo  {journal} {Adv.
  Virus Res.}\ }\textbf {\bibinfo {volume} {16}},\ \bibinfo {pages} {99}
  (\bibinfo {year} {1970})}\BibitemShut {NoStop}%
\bibitem [{\citenamefont {Comas-Garcia}\ \emph {et~al.}(2012)\citenamefont
  {Comas-Garcia}, \citenamefont {Cadena-Nava}, \citenamefont {Rao},
  \citenamefont {Knobler},\ and\ \citenamefont {Gelbart}}]{Comas}%
  \BibitemOpen
  \bibfield  {author} {\bibinfo {author} {\bibfnamefont {M.}~\bibnamefont
  {Comas-Garcia}}, \bibinfo {author} {\bibfnamefont {R.~D.}\ \bibnamefont
  {Cadena-Nava}}, \bibinfo {author} {\bibfnamefont {A.~L.~N.}\ \bibnamefont
  {Rao}}, \bibinfo {author} {\bibfnamefont {C.~M.}\ \bibnamefont {Knobler}}, \
  and\ \bibinfo {author} {\bibfnamefont {W.~M.}\ \bibnamefont {Gelbart}},\
  }\href {\doibase 10.1128/jvi.01695-12} {\bibfield  {journal} {\bibinfo
  {journal} {J. Virol.}\ }\textbf {\bibinfo {volume} {86}},\ \bibinfo {pages}
  {12271} (\bibinfo {year} {2012})}\BibitemShut {NoStop}%
\bibitem [{\citenamefont {Beren}\ \emph {et~al.}(2017)\citenamefont {Beren},
  \citenamefont {Dreesens}, \citenamefont {Liu}, \citenamefont {Knobler},\ and\
  \citenamefont {Gelbart}}]{Beren2017}%
  \BibitemOpen
  \bibfield  {author} {\bibinfo {author} {\bibfnamefont {C.}~\bibnamefont
  {Beren}}, \bibinfo {author} {\bibfnamefont {L.~L.}\ \bibnamefont {Dreesens}},
  \bibinfo {author} {\bibfnamefont {K.~N.}\ \bibnamefont {Liu}}, \bibinfo
  {author} {\bibfnamefont {C.~M.}\ \bibnamefont {Knobler}}, \ and\ \bibinfo
  {author} {\bibfnamefont {W.~M.}\ \bibnamefont {Gelbart}},\ }\href {\doibase
  10.1016/j.bpj.2017.06.038} {\bibfield  {journal} {\bibinfo  {journal}
  {Biophysical Journal}\ }\textbf {\bibinfo {volume} {113}},\ \bibinfo {pages}
  {339} (\bibinfo {year} {2017})}\BibitemShut {NoStop}%
\bibitem [{\citenamefont {Borodavka}\ \emph {et~al.}(2016)\citenamefont
  {Borodavka}, \citenamefont {Singaram}, \citenamefont {Stockley},
  \citenamefont {Gelbart}, \citenamefont {Ben-Shaul},\ and\ \citenamefont
  {Tuma}}]{BORODAVKA2016}%
  \BibitemOpen
  \bibfield  {author} {\bibinfo {author} {\bibfnamefont {A.}~\bibnamefont
  {Borodavka}}, \bibinfo {author} {\bibfnamefont {S.}~\bibnamefont {Singaram}},
  \bibinfo {author} {\bibfnamefont {P.}~\bibnamefont {Stockley}}, \bibinfo
  {author} {\bibfnamefont {W.}~\bibnamefont {Gelbart}}, \bibinfo {author}
  {\bibfnamefont {A.}~\bibnamefont {Ben-Shaul}}, \ and\ \bibinfo {author}
  {\bibfnamefont {R.}~\bibnamefont {Tuma}},\ }\href {\doibase
  https://doi.org/10.1016/j.bpj.2016.10.014} {\bibfield  {journal} {\bibinfo
  {journal} {Biophysical Journal}\ }\textbf {\bibinfo {volume} {111}},\
  \bibinfo {pages} {2077 } (\bibinfo {year} {2016})}\BibitemShut {NoStop}%
\bibitem [{\citenamefont {Hagan}\ and\ \citenamefont
  {Zandi}(2016)}]{Zandi2016}%
  \BibitemOpen
  \bibfield  {author} {\bibinfo {author} {\bibfnamefont {M.~F.}\ \bibnamefont
  {Hagan}}\ and\ \bibinfo {author} {\bibfnamefont {R.}~\bibnamefont {Zandi}},\
  }\href {\doibase 10.1016/j.coviro.2016.02.012} {\bibfield  {journal}
  {\bibinfo  {journal} {Curr. Opin. Virol.}\ }\textbf {\bibinfo {volume}
  {18}},\ \bibinfo {pages} {36} (\bibinfo {year} {2016})}\BibitemShut {NoStop}%
\bibitem [{\citenamefont {Ning}\ \emph {et~al.}(2016)\citenamefont {Ning},
  \citenamefont {Erdemci-Tandogan}, \citenamefont {Yufenyuy}, \citenamefont
  {Wagner}, \citenamefont {Himes}, \citenamefont {Zhao}, \citenamefont {Aiken},
  \citenamefont {Zandi},\ and\ \citenamefont {Zhang}}]{nature2016}%
  \BibitemOpen
  \bibfield  {author} {\bibinfo {author} {\bibfnamefont {J.}~\bibnamefont
  {Ning}}, \bibinfo {author} {\bibfnamefont {G.}~\bibnamefont
  {Erdemci-Tandogan}}, \bibinfo {author} {\bibfnamefont {E.~L.}\ \bibnamefont
  {Yufenyuy}}, \bibinfo {author} {\bibfnamefont {J.}~\bibnamefont {Wagner}},
  \bibinfo {author} {\bibfnamefont {B.~A.}\ \bibnamefont {Himes}}, \bibinfo
  {author} {\bibfnamefont {G.}~\bibnamefont {Zhao}}, \bibinfo {author}
  {\bibfnamefont {C.}~\bibnamefont {Aiken}}, \bibinfo {author} {\bibfnamefont
  {R.}~\bibnamefont {Zandi}}, \ and\ \bibinfo {author} {\bibfnamefont
  {P.}~\bibnamefont {Zhang}},\ }\href {\doibase 10.1038/ncomms13689} {\bibfield
   {journal} {\bibinfo  {journal} {Nature Communications}\ }\textbf {\bibinfo
  {volume} {7}},\ \bibinfo {pages} {13689} (\bibinfo {year}
  {2016})}\BibitemShut {NoStop}%
\bibitem [{\citenamefont {Cadena-Nava}\ \emph {et~al.}(2011)\citenamefont
  {Cadena-Nava}, \citenamefont {Hu}, \citenamefont {Garmann}, \citenamefont
  {Ng}, \citenamefont {Zelikin}, \citenamefont {Knobler},\ and\ \citenamefont
  {Gelbart}}]{Cadena2011}%
  \BibitemOpen
  \bibfield  {author} {\bibinfo {author} {\bibfnamefont {R.~D.}\ \bibnamefont
  {Cadena-Nava}}, \bibinfo {author} {\bibfnamefont {Y.~F.}\ \bibnamefont {Hu}},
  \bibinfo {author} {\bibfnamefont {R.~F.}\ \bibnamefont {Garmann}}, \bibinfo
  {author} {\bibfnamefont {B.}~\bibnamefont {Ng}}, \bibinfo {author}
  {\bibfnamefont {A.~N.}\ \bibnamefont {Zelikin}}, \bibinfo {author}
  {\bibfnamefont {C.~M.}\ \bibnamefont {Knobler}}, \ and\ \bibinfo {author}
  {\bibfnamefont {W.~M.}\ \bibnamefont {Gelbart}},\ }\href {\doibase
  10.1021/jp1094118} {\bibfield  {journal} {\bibinfo  {journal} {J. Phys. Chem.
  B}\ }\textbf {\bibinfo {volume} {115}},\ \bibinfo {pages} {2386} (\bibinfo
  {year} {2011})}\BibitemShut {NoStop}%
\bibitem [{\citenamefont {Perlmutter}\ and\ \citenamefont
  {Hagan}(2015)}]{Haganpackaging}%
  \BibitemOpen
  \bibfield  {author} {\bibinfo {author} {\bibfnamefont {J.~D.}\ \bibnamefont
  {Perlmutter}}\ and\ \bibinfo {author} {\bibfnamefont {M.~F.}\ \bibnamefont
  {Hagan}},\ }\href@noop {} {\bibfield  {journal} {\bibinfo  {journal} {Journal
  of molecular biology}\ }\textbf {\bibinfo {volume} {427}},\ \bibinfo {pages}
  {2451} (\bibinfo {year} {2015})}\BibitemShut {NoStop}%
\bibitem [{\citenamefont {Stockley}\ \emph {et~al.}(2013)\citenamefont
  {Stockley}, \citenamefont {Twarock}, \citenamefont {Bakker}, \citenamefont
  {Barker}, \citenamefont {Borodavka}, \citenamefont {Dykeman}, \citenamefont
  {Ford}, \citenamefont {Pearson}, \citenamefont {Phillips}, \citenamefont
  {Ranson},\ and\ \citenamefont {Tuma}}]{Stockley2013}%
  \BibitemOpen
  \bibfield  {author} {\bibinfo {author} {\bibfnamefont {P.~G.}\ \bibnamefont
  {Stockley}}, \bibinfo {author} {\bibfnamefont {R.}~\bibnamefont {Twarock}},
  \bibinfo {author} {\bibfnamefont {S.~E.}\ \bibnamefont {Bakker}}, \bibinfo
  {author} {\bibfnamefont {A.~M.}\ \bibnamefont {Barker}}, \bibinfo {author}
  {\bibfnamefont {A.}~\bibnamefont {Borodavka}}, \bibinfo {author}
  {\bibfnamefont {E.}~\bibnamefont {Dykeman}}, \bibinfo {author} {\bibfnamefont
  {R.~J.}\ \bibnamefont {Ford}}, \bibinfo {author} {\bibfnamefont {A.~R.}\
  \bibnamefont {Pearson}}, \bibinfo {author} {\bibfnamefont {S.~E.~V.}\
  \bibnamefont {Phillips}}, \bibinfo {author} {\bibfnamefont {N.~A.}\
  \bibnamefont {Ranson}}, \ and\ \bibinfo {author} {\bibfnamefont
  {R.}~\bibnamefont {Tuma}},\ }\href {\doibase 10.1007/s10867-013-9313-0}
  {\bibfield  {journal} {\bibinfo  {journal} {J. Biol. Phys.}\ }\textbf
  {\bibinfo {volume} {39}},\ \bibinfo {pages} {277} (\bibinfo {year}
  {2013})}\BibitemShut {NoStop}%
\bibitem [{\citenamefont {Sun}\ \emph {et~al.}(2007)\citenamefont {Sun},
  \citenamefont {DuFort}, \citenamefont {Daniel}, \citenamefont {Murali},
  \citenamefont {Chen}, \citenamefont {Gopinath}, \citenamefont {Stein},
  \citenamefont {De}, \citenamefont {Rotello}, \citenamefont {Holzenburg},
  \citenamefont {Kao},\ and\ \citenamefont {Dragnea}}]{Sun2007}%
  \BibitemOpen
  \bibfield  {author} {\bibinfo {author} {\bibfnamefont {J.}~\bibnamefont
  {Sun}}, \bibinfo {author} {\bibfnamefont {C.}~\bibnamefont {DuFort}},
  \bibinfo {author} {\bibfnamefont {M.-C.}\ \bibnamefont {Daniel}}, \bibinfo
  {author} {\bibfnamefont {A.}~\bibnamefont {Murali}}, \bibinfo {author}
  {\bibfnamefont {C.}~\bibnamefont {Chen}}, \bibinfo {author} {\bibfnamefont
  {K.}~\bibnamefont {Gopinath}}, \bibinfo {author} {\bibfnamefont
  {B.}~\bibnamefont {Stein}}, \bibinfo {author} {\bibfnamefont
  {M.}~\bibnamefont {De}}, \bibinfo {author} {\bibfnamefont {V.~M.}\
  \bibnamefont {Rotello}}, \bibinfo {author} {\bibfnamefont {A.}~\bibnamefont
  {Holzenburg}}, \bibinfo {author} {\bibfnamefont {C.~C.}\ \bibnamefont {Kao}},
  \ and\ \bibinfo {author} {\bibfnamefont {B.}~\bibnamefont {Dragnea}},\
  }\href@noop {} {\bibfield  {journal} {\bibinfo  {journal} {Proc. Nat. Acad.
  Sci. USA}\ }\textbf {\bibinfo {volume} {104}},\ \bibinfo {pages} {1354}
  (\bibinfo {year} {2007})}\BibitemShut {NoStop}%
\bibitem [{\citenamefont {Li}\ \emph {et~al.}(2017)\citenamefont {Li},
  \citenamefont {Erdemci-Tandogan}, \citenamefont {Wagner}, \citenamefont {{Van
  Der Schoot}},\ and\ \citenamefont {Zandi}}]{Li2017}%
  \BibitemOpen
  \bibfield  {author} {\bibinfo {author} {\bibfnamefont {S.}~\bibnamefont
  {Li}}, \bibinfo {author} {\bibfnamefont {G.}~\bibnamefont
  {Erdemci-Tandogan}}, \bibinfo {author} {\bibfnamefont {J.}~\bibnamefont
  {Wagner}}, \bibinfo {author} {\bibfnamefont {P.}~\bibnamefont {{Van Der
  Schoot}}}, \ and\ \bibinfo {author} {\bibfnamefont {R.}~\bibnamefont
  {Zandi}},\ }\href {\doibase 10.1103/PhysRevE.96.022401} {\bibfield  {journal}
  {\bibinfo  {journal} {Physical Review E}\ }\textbf {\bibinfo {volume} {96}},\
  \bibinfo {pages} {1} (\bibinfo {year} {2017})}\BibitemShut {NoStop}%
\bibitem [{\citenamefont {Zandi}\ \emph {et~al.}(2020)\citenamefont {Zandi},
  \citenamefont {Dragnea}, \citenamefont {Travesset},\ and\ \citenamefont
  {Podgornik}}]{Zandi2020}%
  \BibitemOpen
  \bibfield  {author} {\bibinfo {author} {\bibfnamefont {R.}~\bibnamefont
  {Zandi}}, \bibinfo {author} {\bibfnamefont {B.}~\bibnamefont {Dragnea}},
  \bibinfo {author} {\bibfnamefont {A.}~\bibnamefont {Travesset}}, \ and\
  \bibinfo {author} {\bibfnamefont {R.}~\bibnamefont {Podgornik}},\ }\href
  {\doibase 10.1016/J.PHYSREP.2019.12.005} {\bibfield  {journal} {\bibinfo
  {journal} {Phys. Rep.}\ }\textbf {\bibinfo {volume} {847}},\ \bibinfo {pages}
  {1} (\bibinfo {year} {2020})}\BibitemShut {NoStop}%
\bibitem [{\citenamefont {Sivanandam}\ \emph {et~al.}(2016)\citenamefont
  {Sivanandam}, \citenamefont {Mathews}, \citenamefont {Garmann}, \citenamefont
  {Erdemci-Tandogan}, \citenamefont {Zandi},\ and\ \citenamefont
  {Rao}}]{Venky2016}%
  \BibitemOpen
  \bibfield  {author} {\bibinfo {author} {\bibfnamefont {V.}~\bibnamefont
  {Sivanandam}}, \bibinfo {author} {\bibfnamefont {D.}~\bibnamefont {Mathews}},
  \bibinfo {author} {\bibfnamefont {R.}~\bibnamefont {Garmann}}, \bibinfo
  {author} {\bibfnamefont {G.}~\bibnamefont {Erdemci-Tandogan}}, \bibinfo
  {author} {\bibfnamefont {R.}~\bibnamefont {Zandi}}, \ and\ \bibinfo {author}
  {\bibfnamefont {A.~L.~N.}\ \bibnamefont {Rao}},\ }\href
  {http://dx.doi.org/10.1038/srep26328 http://10.1038/srep26328} {\bibfield
  {journal} {\bibinfo  {journal} {Scientific Reports}\ }\textbf {\bibinfo
  {volume} {6}},\ \bibinfo {pages} {26328} (\bibinfo {year}
  {2016})}\BibitemShut {NoStop}%
\bibitem [{\citenamefont {Ni}\ \emph {et~al.}(2012)\citenamefont {Ni},
  \citenamefont {Wang}, \citenamefont {Ma}, \citenamefont {Das}, \citenamefont
  {Sokol}, \citenamefont {Chiu}, \citenamefont {Dragnea}, \citenamefont
  {Hagan},\ and\ \citenamefont {Kao}}]{Bogdan}%
  \BibitemOpen
  \bibfield  {author} {\bibinfo {author} {\bibfnamefont {P.}~\bibnamefont
  {Ni}}, \bibinfo {author} {\bibfnamefont {Z.}~\bibnamefont {Wang}}, \bibinfo
  {author} {\bibfnamefont {X.}~\bibnamefont {Ma}}, \bibinfo {author}
  {\bibfnamefont {N.~C.}\ \bibnamefont {Das}}, \bibinfo {author} {\bibfnamefont
  {P.}~\bibnamefont {Sokol}}, \bibinfo {author} {\bibfnamefont
  {W.}~\bibnamefont {Chiu}}, \bibinfo {author} {\bibfnamefont {B.}~\bibnamefont
  {Dragnea}}, \bibinfo {author} {\bibfnamefont {M.}~\bibnamefont {Hagan}}, \
  and\ \bibinfo {author} {\bibfnamefont {C.~C.}\ \bibnamefont {Kao}},\ }\href
  {\doibase 10.1016/j.jmb.2012.03.023} {\bibfield  {journal} {\bibinfo
  {journal} {J. Mol. Biol.}\ }\textbf {\bibinfo {volume} {419}},\ \bibinfo
  {pages} {284} (\bibinfo {year} {2012})}\BibitemShut {NoStop}%
\bibitem [{\citenamefont {Belyi}\ and\ \citenamefont
  {Muthukumar}(2006)}]{Belyi2006}%
  \BibitemOpen
  \bibfield  {author} {\bibinfo {author} {\bibfnamefont {V.~A.}\ \bibnamefont
  {Belyi}}\ and\ \bibinfo {author} {\bibfnamefont {M.}~\bibnamefont
  {Muthukumar}},\ }\href {\doibase 10.1073/pnas.0608311103} {\bibfield
  {journal} {\bibinfo  {journal} {PNAS}\ }\textbf {\bibinfo {volume} {103}},\
  \bibinfo {pages} {17174} (\bibinfo {year} {2006})}\BibitemShut {NoStop}%
\bibitem [{\citenamefont {Tao}\ \emph {et~al.}(2008)\citenamefont {Tao},
  \citenamefont {Rui},\ and\ \citenamefont {Shklovskii}}]{Shklovskii}%
  \BibitemOpen
  \bibfield  {author} {\bibinfo {author} {\bibfnamefont {H.}~\bibnamefont
  {Tao}}, \bibinfo {author} {\bibfnamefont {Z.}~\bibnamefont {Rui}}, \ and\
  \bibinfo {author} {\bibfnamefont {B.~I.}\ \bibnamefont {Shklovskii}},\ }\href
  {\doibase 10.1016/j.physa.2008.01.010} {\bibfield  {journal} {\bibinfo
  {journal} {Physica A}\ }\textbf {\bibinfo {volume} {387}},\ \bibinfo {pages}
  {3059} (\bibinfo {year} {2008})}\BibitemShut {NoStop}%
\bibitem [{\citenamefont {Zeng}\ \emph {et~al.}(2017)\citenamefont {Zeng},
  \citenamefont {Hernando-P{\'{e}}rez}, \citenamefont {Dragnea}, \citenamefont
  {Ma}, \citenamefont {van~der Schoot},\ and\ \citenamefont
  {Zandi}}]{Zeng2017a}%
  \BibitemOpen
  \bibfield  {author} {\bibinfo {author} {\bibfnamefont {C.}~\bibnamefont
  {Zeng}}, \bibinfo {author} {\bibfnamefont {M.}~\bibnamefont
  {Hernando-P{\'{e}}rez}}, \bibinfo {author} {\bibfnamefont {B.}~\bibnamefont
  {Dragnea}}, \bibinfo {author} {\bibfnamefont {X.}~\bibnamefont {Ma}},
  \bibinfo {author} {\bibfnamefont {P.}~\bibnamefont {van~der Schoot}}, \ and\
  \bibinfo {author} {\bibfnamefont {R.}~\bibnamefont {Zandi}},\ }\href@noop {}
  {\bibfield  {journal} {\bibinfo  {journal} {Phys. Rev. Lett.}\ }\textbf
  {\bibinfo {volume} {119}},\ \bibinfo {pages} {038102} (\bibinfo {year}
  {2017})}\BibitemShut {NoStop}%
\bibitem [{\citenamefont {Caspar}\ and\ \citenamefont
  {Klug}(1962)}]{CASPAR1962}%
  \BibitemOpen
  \bibfield  {author} {\bibinfo {author} {\bibfnamefont {D.~L.}\ \bibnamefont
  {Caspar}}\ and\ \bibinfo {author} {\bibfnamefont {A.}~\bibnamefont {Klug}},\
  }\href {\doibase doi:10.1101/SQB.1962.027.001.005} {\bibfield  {journal}
  {\bibinfo  {journal} {Cold Spring Harbor Symp. Quant. Biol.}\ }\textbf
  {\bibinfo {volume} {27}},\ \bibinfo {pages} {1} (\bibinfo {year}
  {1962})}\BibitemShut {NoStop}%
\bibitem [{\citenamefont {Gopal}\ \emph {et~al.}(2014)\citenamefont {Gopal},
  \citenamefont {D.E.}, \citenamefont {A.M.}, \citenamefont {A}, \citenamefont
  {ALN}, \citenamefont {Knobler}, \citenamefont {Gelbart},\ and\ \citenamefont
  {Ben-Shaul}}]{Gopal2014}%
  \BibitemOpen
  \bibfield  {author} {\bibinfo {author} {\bibfnamefont {A.}~\bibnamefont
  {Gopal}}, \bibinfo {author} {\bibfnamefont {E.}~\bibnamefont {D.E.}},
  \bibinfo {author} {\bibfnamefont {Y.}~\bibnamefont {A.M.}}, \bibinfo {author}
  {\bibfnamefont {B.-S.}\ \bibnamefont {A}}, \bibinfo {author} {\bibfnamefont
  {R.}~\bibnamefont {ALN}}, \bibinfo {author} {\bibfnamefont {C.~M.}\
  \bibnamefont {Knobler}}, \bibinfo {author} {\bibfnamefont {W.~M.}\
  \bibnamefont {Gelbart}}, \ and\ \bibinfo {author} {\bibfnamefont
  {A.}~\bibnamefont {Ben-Shaul}},\ }\href@noop {} {\bibfield  {journal}
  {\bibinfo  {journal} {PLoS ONE}\ }\textbf {\bibinfo {volume} {9}},\ \bibinfo
  {pages} {e105875} (\bibinfo {year} {2014})}\BibitemShut {NoStop}%
\bibitem [{\citenamefont {Perlmutter}\ \emph {et~al.}(2013)\citenamefont
  {Perlmutter}, \citenamefont {Qiao},\ and\ \citenamefont {Hagan}}]{elife}%
  \BibitemOpen
  \bibfield  {author} {\bibinfo {author} {\bibfnamefont {J.~D.}\ \bibnamefont
  {Perlmutter}}, \bibinfo {author} {\bibfnamefont {C.}~\bibnamefont {Qiao}}, \
  and\ \bibinfo {author} {\bibfnamefont {M.~F.}\ \bibnamefont {Hagan}},\ }\href
  {\doibase 10.7554/eLife.00632} {\bibfield  {journal} {\bibinfo  {journal}
  {eLife}\ }\textbf {\bibinfo {volume} {2}} (\bibinfo {year} {2013}),\
  10.7554/eLife.00632}\BibitemShut {NoStop}%
\bibitem [{\citenamefont {van~der Schoot}\ and\ \citenamefont
  {Zandi}(2013)}]{Paul:13a}%
  \BibitemOpen
  \bibfield  {author} {\bibinfo {author} {\bibfnamefont {P.}~\bibnamefont
  {van~der Schoot}}\ and\ \bibinfo {author} {\bibfnamefont {R.}~\bibnamefont
  {Zandi}},\ }\href {\doibase 10.1007/s10867-013-9307-y} {\bibfield  {journal}
  {\bibinfo  {journal} {J. Biol. Phys.}\ }\textbf {\bibinfo {volume} {39}},\
  \bibinfo {pages} {289} (\bibinfo {year} {2013})}\BibitemShut {NoStop}%
\bibitem [{\citenamefont {de~Gennes}(1979)}]{deGennes1979}%
  \BibitemOpen
  \bibfield  {author} {\bibinfo {author} {\bibfnamefont {P.-G.}\ \bibnamefont
  {de~Gennes}},\ }\href@noop {} {\emph {\bibinfo {title} {Scaling concepts in
  polymer physics}}}\ (\bibinfo  {publisher} {Cornell University Press},\
  \bibinfo {address} {Ithaca, New York},\ \bibinfo {year} {1979})\BibitemShut
  {NoStop}%
\bibitem [{\citenamefont {Li}\ \emph {et~al.}(2018)\citenamefont {Li},
  \citenamefont {Orland},\ and\ \citenamefont {Zandi}}]{Li_2018}%
  \BibitemOpen
  \bibfield  {author} {\bibinfo {author} {\bibfnamefont {S.}~\bibnamefont
  {Li}}, \bibinfo {author} {\bibfnamefont {H.}~\bibnamefont {Orland}}, \ and\
  \bibinfo {author} {\bibfnamefont {R.}~\bibnamefont {Zandi}},\ }\href
  {\doibase 10.1088/1361-648x/aab0c6} {\bibfield  {journal} {\bibinfo
  {journal} {Journal of Physics: Condensed Matter}\ }\textbf {\bibinfo {volume}
  {30}},\ \bibinfo {pages} {144002} (\bibinfo {year} {2018})}\BibitemShut
  {NoStop}%
\bibitem [{\citenamefont {Borukhov}\ \emph {et~al.}(1998)\citenamefont
  {Borukhov}, \citenamefont {Andelman},\ and\ \citenamefont
  {Orland}}]{Borukhov}%
  \BibitemOpen
  \bibfield  {author} {\bibinfo {author} {\bibfnamefont {I.}~\bibnamefont
  {Borukhov}}, \bibinfo {author} {\bibfnamefont {D.}~\bibnamefont {Andelman}},
  \ and\ \bibinfo {author} {\bibfnamefont {H.}~\bibnamefont {Orland}},\ }\href
  {\doibase 10.1007/s100510050513} {\bibfield  {journal} {\bibinfo  {journal}
  {Euro. Phys. J. B}\ }\textbf {\bibinfo {volume} {5}},\ \bibinfo {pages} {869}
  (\bibinfo {year} {1998})}\BibitemShut {NoStop}%
\bibitem [{\citenamefont {Siber}\ and\ \citenamefont
  {Podgornik}(2008)}]{Siber2008}%
  \BibitemOpen
  \bibfield  {author} {\bibinfo {author} {\bibfnamefont {A.}~\bibnamefont
  {Siber}}\ and\ \bibinfo {author} {\bibfnamefont {R.}~\bibnamefont
  {Podgornik}},\ }\href {\doibase 10.1103/PhysRevE.78.051915} {\bibfield
  {journal} {\bibinfo  {journal} {Phys. Rev. E}\ }\textbf {\bibinfo {volume}
  {78}},\ \bibinfo {pages} {051915} (\bibinfo {year} {2008})}\BibitemShut
  {NoStop}%
\bibitem [{\citenamefont {Erdemci-Tandogan}\ \emph {et~al.}(2014)\citenamefont
  {Erdemci-Tandogan}, \citenamefont {Wagner}, \citenamefont {van~der Schoot},
  \citenamefont {Podgornik},\ and\ \citenamefont {Zandi}}]{Gonca2014}%
  \BibitemOpen
  \bibfield  {author} {\bibinfo {author} {\bibfnamefont {G.}~\bibnamefont
  {Erdemci-Tandogan}}, \bibinfo {author} {\bibfnamefont {J.}~\bibnamefont
  {Wagner}}, \bibinfo {author} {\bibfnamefont {P.}~\bibnamefont {van~der
  Schoot}}, \bibinfo {author} {\bibfnamefont {R.}~\bibnamefont {Podgornik}}, \
  and\ \bibinfo {author} {\bibfnamefont {R.}~\bibnamefont {Zandi}},\ }\href
  {\doibase 10.1103/PhysRevE.89.032707} {\bibfield  {journal} {\bibinfo
  {journal} {Phys. Rev. E}\ }\textbf {\bibinfo {volume} {89}},\ \bibinfo
  {pages} {032707} (\bibinfo {year} {2014})}\BibitemShut {NoStop}%
\bibitem [{\citenamefont {Erdemci-Tandogan}\ \emph {et~al.}(2016)\citenamefont
  {Erdemci-Tandogan}, \citenamefont {Wagner}, \citenamefont {van~der Schoot},
  \citenamefont {Podgornik},\ and\ \citenamefont {Zandi}}]{Gonca2016}%
  \BibitemOpen
  \bibfield  {author} {\bibinfo {author} {\bibfnamefont {G.}~\bibnamefont
  {Erdemci-Tandogan}}, \bibinfo {author} {\bibfnamefont {J.}~\bibnamefont
  {Wagner}}, \bibinfo {author} {\bibfnamefont {P.}~\bibnamefont {van~der
  Schoot}}, \bibinfo {author} {\bibfnamefont {R.}~\bibnamefont {Podgornik}}, \
  and\ \bibinfo {author} {\bibfnamefont {R.}~\bibnamefont {Zandi}},\
  }\href@noop {} {\bibfield  {journal} {\bibinfo  {journal} {Phys. Rev. E}\
  }\textbf {\bibinfo {volume} {94}},\ \bibinfo {pages} {022408} (\bibinfo
  {year} {2016})}\BibitemShut {NoStop}%
\bibitem [{\citenamefont {Janssen}\ \emph {et~al.}(2014)\citenamefont
  {Janssen}, \citenamefont {H\"artel},\ and\ \citenamefont {van
  Roij}}]{Janssen2014}%
  \BibitemOpen
  \bibfield  {author} {\bibinfo {author} {\bibfnamefont {M.}~\bibnamefont
  {Janssen}}, \bibinfo {author} {\bibfnamefont {A.}~\bibnamefont {H\"artel}}, \
  and\ \bibinfo {author} {\bibfnamefont {R.}~\bibnamefont {van Roij}},\ }\href
  {\doibase 10.1103/PhysRevLett.113.268501} {\bibfield  {journal} {\bibinfo
  {journal} {Phys. Rev. Lett.}\ }\textbf {\bibinfo {volume} {113}},\ \bibinfo
  {pages} {268501} (\bibinfo {year} {2014})}\BibitemShut {NoStop}%
\bibitem [{\citenamefont {Wagner}\ \emph {et~al.}(2015)\citenamefont {Wagner},
  \citenamefont {Erdemci-Tandogan},\ and\ \citenamefont
  {Zandi}}]{adsorption2015}%
  \BibitemOpen
  \bibfield  {author} {\bibinfo {author} {\bibfnamefont {J.}~\bibnamefont
  {Wagner}}, \bibinfo {author} {\bibfnamefont {G.}~\bibnamefont
  {Erdemci-Tandogan}}, \ and\ \bibinfo {author} {\bibfnamefont
  {R.}~\bibnamefont {Zandi}},\ }\href {\doibase 10.1088/0953-8984/27/49/495101}
  {\bibfield  {journal} {\bibinfo  {journal} {J. Phys.:Condens. Matter}\
  }\textbf {\bibinfo {volume} {27}},\ \bibinfo {pages} {495101} (\bibinfo
  {year} {2015})}\BibitemShut {NoStop}%
\bibitem [{\citenamefont {Borukhov}\ \emph {et~al.}(1995)\citenamefont
  {Borukhov}, \citenamefont {Andelman},\ and\ \citenamefont
  {Orland}}]{Borukhov1}%
  \BibitemOpen
  \bibfield  {author} {\bibinfo {author} {\bibfnamefont {I.}~\bibnamefont
  {Borukhov}}, \bibinfo {author} {\bibfnamefont {D.}~\bibnamefont {Andelman}},
  \ and\ \bibinfo {author} {\bibfnamefont {H.}~\bibnamefont {Orland}},\ }\href
  {\doibase 10.1209/0295-5075/32/6/007} {\bibfield  {journal} {\bibinfo
  {journal} {Europhys. Lett.}\ }\textbf {\bibinfo {volume} {32}},\ \bibinfo
  {pages} {499} (\bibinfo {year} {1995})}\BibitemShut {NoStop}%
\bibitem [{\citenamefont {Shafir}\ \emph {et~al.}(2003)\citenamefont {Shafir},
  \citenamefont {Andelman},\ and\ \citenamefont {Netz}}]{Shafir}%
  \BibitemOpen
  \bibfield  {author} {\bibinfo {author} {\bibfnamefont {A.}~\bibnamefont
  {Shafir}}, \bibinfo {author} {\bibfnamefont {D.}~\bibnamefont {Andelman}}, \
  and\ \bibinfo {author} {\bibfnamefont {R.~R.}\ \bibnamefont {Netz}},\ }\href
  {\doibase 10.1063/1.1580798} {\bibfield  {journal} {\bibinfo  {journal} {J.
  Chem. Phys.}\ }\textbf {\bibinfo {volume} {119}},\ \bibinfo {pages} {2355}
  (\bibinfo {year} {2003})}\BibitemShut {NoStop}%
\bibitem [{\citenamefont {Siber}\ \emph {et~al.}(2012)\citenamefont {Siber},
  \citenamefont {Bozic},\ and\ \citenamefont {Podgornik}}]{Siber-nonspecific}%
  \BibitemOpen
  \bibfield  {author} {\bibinfo {author} {\bibfnamefont {A.}~\bibnamefont
  {Siber}}, \bibinfo {author} {\bibfnamefont {A.~L.}\ \bibnamefont {Bozic}}, \
  and\ \bibinfo {author} {\bibfnamefont {R.}~\bibnamefont {Podgornik}},\ }\href
  {\doibase 10.1039/c1cp22756d} {\bibfield  {journal} {\bibinfo  {journal}
  {Phys. Chem. Chem. Phys.}\ }\textbf {\bibinfo {volume} {14}},\ \bibinfo
  {pages} {3746} (\bibinfo {year} {2012})},\ \Eprint
  {http://arxiv.org/abs/5905} {arXiv:1108:5905} \BibitemShut {NoStop}%
\bibitem [{\citenamefont {Lubensky}\ and\ \citenamefont
  {Isaacson}(1979)}]{Lubensky}%
  \BibitemOpen
  \bibfield  {author} {\bibinfo {author} {\bibfnamefont {T.~C.}\ \bibnamefont
  {Lubensky}}\ and\ \bibinfo {author} {\bibfnamefont {J.}~\bibnamefont
  {Isaacson}},\ }\href {\doibase 10.1103/PhysRevA.20.2130} {\bibfield
  {journal} {\bibinfo  {journal} {Phys. Rev. A}\ }\textbf {\bibinfo {volume}
  {20}},\ \bibinfo {pages} {2130} (\bibinfo {year} {1979})}\BibitemShut
  {NoStop}%
\bibitem [{\citenamefont {Lee}\ and\ \citenamefont
  {Nguyen}(2008)}]{Lee-Nguyen}%
  \BibitemOpen
  \bibfield  {author} {\bibinfo {author} {\bibfnamefont {S.~I.}\ \bibnamefont
  {Lee}}\ and\ \bibinfo {author} {\bibfnamefont {T.~T.}\ \bibnamefont
  {Nguyen}},\ }\href {\doibase 10.1103/PhysRevLett.100.198102} {\bibfield
  {journal} {\bibinfo  {journal} {Phys. Rev. Lett.}\ }\textbf {\bibinfo
  {volume} {100}},\ \bibinfo {pages} {198102} (\bibinfo {year}
  {2008})}\BibitemShut {NoStop}%
\bibitem [{\citenamefont {Elleuch}\ \emph {et~al.}(1995)\citenamefont
  {Elleuch}, \citenamefont {Lequeux},\ and\ \citenamefont {Pfeuty}}]{Elleuch}%
  \BibitemOpen
  \bibfield  {author} {\bibinfo {author} {\bibfnamefont {K.}~\bibnamefont
  {Elleuch}}, \bibinfo {author} {\bibfnamefont {F.}~\bibnamefont {Lequeux}}, \
  and\ \bibinfo {author} {\bibfnamefont {P.}~\bibnamefont {Pfeuty}},\ }\href
  {\doibase 10.1051/jp1:1995140} {\bibfield  {journal} {\bibinfo  {journal} {J.
  Phys. I France}\ }\textbf {\bibinfo {volume} {5}},\ \bibinfo {pages} {465}
  (\bibinfo {year} {1995})}\BibitemShut {NoStop}%
\bibitem [{\citenamefont {de~Gennes}(1982)}]{deGennes}%
  \BibitemOpen
  \bibfield  {author} {\bibinfo {author} {\bibfnamefont {P.-G.}\ \bibnamefont
  {de~Gennes}},\ }\href@noop {} {\bibfield  {journal} {\bibinfo  {journal}
  {Macromolecules}\ }\textbf {\bibinfo {volume} {15}},\ \bibinfo {pages} {492}
  (\bibinfo {year} {1982})}\BibitemShut {NoStop}%
\bibitem [{\citenamefont {Ji}\ and\ \citenamefont {Hone}(1988)}]{Hone}%
  \BibitemOpen
  \bibfield  {author} {\bibinfo {author} {\bibfnamefont {H.}~\bibnamefont
  {Ji}}\ and\ \bibinfo {author} {\bibfnamefont {D.}~\bibnamefont {Hone}},\
  }\href {\doibase 10.1021/ma00186a049} {\bibfield  {journal} {\bibinfo
  {journal} {Macromolecules}\ }\textbf {\bibinfo {volume} {21}},\ \bibinfo
  {pages} {2600} (\bibinfo {year} {1988})}\BibitemShut {NoStop}%
\bibitem [{\citenamefont {Bangerth}\ \emph {et~al.}(2007)\citenamefont
  {Bangerth}, \citenamefont {Hartmann},\ and\ \citenamefont
  {Kanschat}}]{BangerthHartmannKanschat2007}%
  \BibitemOpen
  \bibfield  {author} {\bibinfo {author} {\bibfnamefont {W.}~\bibnamefont
  {Bangerth}}, \bibinfo {author} {\bibfnamefont {R.}~\bibnamefont {Hartmann}},
  \ and\ \bibinfo {author} {\bibfnamefont {G.}~\bibnamefont {Kanschat}},\
  }\href@noop {} {\bibfield  {journal} {\bibinfo  {journal} {ACM Trans. Math.
  Softw.}\ }\textbf {\bibinfo {volume} {33}},\ \bibinfo {pages} {24/1}
  (\bibinfo {year} {2007})}\BibitemShut {NoStop}%
\bibitem [{\citenamefont {Bathe}(1996)}]{bathe1996finite}%
  \BibitemOpen
  \bibfield  {author} {\bibinfo {author} {\bibfnamefont {K.}~\bibnamefont
  {Bathe}},\ }\href {https://books.google.com/books?id=AJBRAAAAMAAJ} {\emph
  {\bibinfo {title} {Finite Element Procedures}}},\ \bibinfo {series} {Finite
  Element Procedures}\ No.\ \bibinfo {number} {pt. 2}\ (\bibinfo  {publisher}
  {Prentice Hall},\ \bibinfo {address} {New Jersey},\ \bibinfo {year}
  {1996})\BibitemShut {NoStop}%
\bibitem [{\citenamefont {Nocedal}\ and\ \citenamefont
  {Wright}(2006)}]{NoceWrig06}%
  \BibitemOpen
  \bibfield  {author} {\bibinfo {author} {\bibfnamefont {J.}~\bibnamefont
  {Nocedal}}\ and\ \bibinfo {author} {\bibfnamefont {S.~J.}\ \bibnamefont
  {Wright}},\ }\href@noop {} {\emph {\bibinfo {title} {Numerical
  Optimization}}},\ \bibinfo {edition} {2nd}\ ed.\ (\bibinfo  {publisher}
  {Springer},\ \bibinfo {address} {New York, NY},\ \bibinfo {year}
  {2006})\BibitemShut {NoStop}%
\bibitem [{\citenamefont {Erdemci-Tandogan}\ \emph {et~al.}(2017)\citenamefont
  {Erdemci-Tandogan}, \citenamefont {Orland},\ and\ \citenamefont
  {Zandi}}]{GoncaPRL2017}%
  \BibitemOpen
  \bibfield  {author} {\bibinfo {author} {\bibfnamefont {G.}~\bibnamefont
  {Erdemci-Tandogan}}, \bibinfo {author} {\bibfnamefont {H.}~\bibnamefont
  {Orland}}, \ and\ \bibinfo {author} {\bibfnamefont {R.}~\bibnamefont
  {Zandi}},\ }\href {\doibase 10.1103/PhysRevLett.119.188102} {\bibfield
  {journal} {\bibinfo  {journal} {Phys. Rev. Lett.}\ }\textbf {\bibinfo
  {volume} {119}},\ \bibinfo {pages} {188102} (\bibinfo {year}
  {2017})}\BibitemShut {NoStop}%
\bibitem [{\citenamefont {Patel}\ \emph {et~al.}(2015)\citenamefont {Patel},
  \citenamefont {Dykeman}, \citenamefont {Coutts}, \citenamefont {Lomonossoff},
  \citenamefont {Rowlands}, \citenamefont {Phillips}, \citenamefont {Ranson},
  \citenamefont {Twarock}, \citenamefont {Tuma},\ and\ \citenamefont
  {Stockley}}]{Patel2014}%
  \BibitemOpen
  \bibfield  {author} {\bibinfo {author} {\bibfnamefont {N.}~\bibnamefont
  {Patel}}, \bibinfo {author} {\bibfnamefont {E.~C.}\ \bibnamefont {Dykeman}},
  \bibinfo {author} {\bibfnamefont {R.~H.~A.}\ \bibnamefont {Coutts}}, \bibinfo
  {author} {\bibfnamefont {G.~P.}\ \bibnamefont {Lomonossoff}}, \bibinfo
  {author} {\bibfnamefont {D.~J.}\ \bibnamefont {Rowlands}}, \bibinfo {author}
  {\bibfnamefont {S.~E.~V.}\ \bibnamefont {Phillips}}, \bibinfo {author}
  {\bibfnamefont {N.}~\bibnamefont {Ranson}}, \bibinfo {author} {\bibfnamefont
  {R.}~\bibnamefont {Twarock}}, \bibinfo {author} {\bibfnamefont
  {R.}~\bibnamefont {Tuma}}, \ and\ \bibinfo {author} {\bibfnamefont {P.~G.}\
  \bibnamefont {Stockley}},\ }\href {\doibase 10.1073/pnas.1420812112}
  {\bibfield  {journal} {\bibinfo  {journal} {Proceedings of the National
  Academy of Sciences}\ }\textbf {\bibinfo {volume} {112}},\ \bibinfo {pages}
  {2227} (\bibinfo {year} {2015})},\ \Eprint
  {http://arxiv.org/abs/http://www.pnas.org/content/112/7/2227.full.pdf}
  {http://www.pnas.org/content/112/7/2227.full.pdf} \BibitemShut {NoStop}%
\end{thebibliography}
%

\end{document}